\documentclass[12pt]{article}
\usepackage[paper=a4paper,dvips,top=2.5cm,left=2.5cm,right=2.5cm,
    foot=2cm,bottom=2.5cm]{geometry}
\usepackage{hyperref}
\hypersetup{
    colorlinks=true,
    linkcolor=blue,
    filecolor=blue,      
    urlcolor=blue,
    citecolor=blue,
    pdftitle={Rapid prediction of cardiac activation in the left ventricle with geometric deep learning: a step towards cardiac resynchronization therapy planning},
    pdfpagemode=FullScreen,
    }
\usepackage{amsmath,amssymb}

\DeclareMathOperator*{\argmin}{arg\,min}
\usepackage{graphicx}
\usepackage{appendix}
\usepackage{blindtext}
\usepackage{subfiles}
\usepackage{multirow}
\usepackage{booktabs}
\usepackage{tabularx}
\usepackage{siunitx}
\usepackage[font=footnotesize]{caption}
\usepackage[usenames,dvipsnames]{color}
\usepackage{authblk}

\usepackage{rotating}
\usepackage{makecell}
\usepackage{array}
\usepackage{xcolor}
\usepackage{floatrow}
\usepackage{multicol}
\usepackage{bm}
\usepackage{subcaption}
\usepackage[backend=biber,
            style=nature,
            sorting=none,
            maxbibnames=5,
            minbibnames=3,
            doi=false,
            isbn=false,
            url=false,
            eprint=false,
            backref=true]{biblatex}
\AtBeginBibliography{\small}

\DeclareMathAlphabet{\msfsl}{T1}{cmtt}{m}{it}
\title{Rapid prediction of cardiac activation in the left ventricle with geometric deep learning: a step towards cardiac resynchronization therapy planning}

\author[1,2]{Ehsan Naghavi}
\author[1]{Haifeng Wang}
\author[1]{Vahid Ziaei-Rad}
\author[2,3]{Julius Guccione}
\author[2,3]{Ghassan Kassab}
\author[4]{Vishnu Boddeti}
\author[1]{Seungik Baek}
\author[1]{Lik-Chuan Lee\thanks{Corresponding author: lclee@egr.msu.edu}}
\affil[1]{Dept. of Mechanical Engineering, Michigan State University, East Lansing, MI}
\affil[2]{3DT Holdings, LLC, San Diego, CA}
\affil[3]{California Medical Innovations Institute, San Diego, CA}
\affil[4]{Dept. of Computer Science and Engineering, Michigan State University, East Lansing, MI}
\date{}

\begin{document}
\maketitle

\begin{abstract}  
    Cardiac resynchronization therapy (CRT) is a common intervention for patients with dyssynchronous heart failure, yet approximately one-third of recipients fail to respond, partly due to suboptimal lead placement. Identifying optimal pacing sites remains challenging, largely due to patient-specific anatomical variability and limitations of current individualized planning strategies. In a step toward an \textit{in-silico} approach, we develop two geometric deep learning models, based on graph neural network (GNN) and geometry-informed neural operator (GINO), to predict activation time maps on left ventricular (LV) geometries in real time. Trained on a large dataset generated from finite-element simulations spanning a wide range of synthetic LV shapes, pacing site configurations, and tissue conductivities, the GINO model outperforms the GNN on synthetic cases (1.38\% vs 2.44\% error), while both demonstrate comparable performance on real-world LV geometries (GINO: 4.79\% vs GNN: 4.07\%). Using the trained models, we develop a workflow to identify an optimal pacing site on the LV from a given activation time map and show that both models can robustly recover ground-truth subject-specific parameters from noisy inputs. In conjunction with an interactive web-based interface (\url{https://dcsim.egr.msu.edu/}), this study shows potential and motivates future extension toward a clinical decision-support tool for personalized pre-procedural CRT optimization.
    \hfill

    \vspace{0.5cm}

    \noindent \textit{\textbf{Keywords:}} Cardiac resynchronization therapy, Geometric deep learning, Neural operators, Graph neural networks, Cardiac electrophysiology

    \hfill

\end{abstract}

\section{Introduction}
In patients with dyssynchronous heart failure, cardiac resynchronization therapy (CRT) is an established treatment to restore synchronized ventricular activation patterns~\autocite{Glikson2022, Nisha2017, McAlister2007, WilliamT2002}. Despite its efficacy, approximately one-third of CRT recipients do not experience optimal outcomes and are classified as non-responders~\autocite{Glikson2022, Ypenburg2009}. Contributing factors of CRT non-response include the heterogeneity of left ventricular (LV) anatomy and challenges associated with individualized lead placement strategies~\autocite{Wijesuriya2023, WOUTERS20211024, Bogaard2010, Wouter2015, Jagmeet2011}. Although the importance of pacing site selection is emphasized in numerous experimental and clinical studies~\autocite{Svendsen2012, Jagmeet2011, SAXON2009, Helm2007, Butter2000}, its optimization in clinical practice remains difficult~\autocite{FAN2022105050}.

To address this issue, physics-based computational models based on differential equations have been developed to provide insights into CRT mechanisms and assist in therapy optimization~\autocite{Craine2024, ALBATAT2021104159, Carpio2019, AWCLee2018, Villongco2016, Pluijmert2016, Huntjens2014, NIEDERER2012, CONSTANTINO2012372, KERCKHOFFS2009362}. While these models can provide high-fidelity predictions, their reliance on computationally expensive numerical methods limits their practical use in time-sensitive clinical scenarios. To overcome this limitation, deep learning (DL) methodologies are increasingly used to develop surrogate models capable of rapid predictions with reasonable accuracy~\autocite{NAGHAVI2024102995, PILIA2023102619, Sacks2022, ruiz2022physics, BUOSO2021102066, KONG2021102222, Peng2021, Rasheed2020, Dabiri2020, Costabal2020, Dabiri2019, Alber2019, maher2019accelerating}. DL models specifically for optimizing pacing site selection in CRT, however, remain scarce. Most machine learning studies have primarily focused on classifying treatment responses based on clinical characteristics (e.g., ischemia versus non-ischemia)~\autocite{CHANG2024100803, Odigwe2022, Wouters2022, Tokodi2020, Feeny2019, Kalscheur2018}, rather than modeling and optimizing the effects of various pacing sites to enhance therapeutic outcomes.

Accommodating arbitrary patient-specific heart geometries is a major challenge for developing DL surrogates. Convolutional neural network (CNN)-based methods are restricted to structured grids~\autocite{Fresca2020}, and approaches tailored to single geometries necessitate retraining for each new patient~\autocite{REGAZZONI2022114825, Giffard2019}. Reduced-order models~\autocite{Fresca2021, Pagani2021}, though offering some flexibility, often fail to capture localized anatomical features crucial for precise predictions, thereby restricting the clinical utility of DL-based CRT planning tools.

Graph neural networks (GNNs) have emerged as a promising approach for learning simulations in irregular domains~\autocite{LearningtoSimCompPhysGN2020}. By representing geometries as graph-structured data composed of nodes and edges, GNNs naturally handle arbitrary geometries and have been applied in cardiovascular research~\autocite{Morales2021, PIGNN2024, IVANTSITS2024109154, DALTON2023116351, PEGOLOTTI2024107676}. Concurrently, neural operators have recently gained attention for their ability to learn mappings between infinite-dimensional function spaces, in contrast to conventional neural networks that typically map between finite-dimensional vector spaces~\autocite{kovachki2023neural}. This capability offers a principled approach to modeling solution operators of differential equations, making neural operators particularly suitable for scientific modeling tasks on continuous domains~\autocite{Azizzadenesheli2024}, such as in cardiovascular simulations. Despite their potential, neural operators have only recently been applied to cardiovascular problems~\autocite{DIMON2024}, in part because implementation challenges and their initial development for regular input grids~\autocite{li2021Fourier} limited their applicability to anatomical domains.

Here, we develop and compare two physics-based DL models to predict cardiac activation patterns in the LV as a step toward CRT planning. The first model utilizes a GNN architecture (Graph-UNet) ~\autocite{gao2019graphunets}, whereas the second employs geometry-informed neural operator (GINO)~\autocite{li2023geometryinformed}, a neural operator variant designed to handle arbitrary input geometries. Both models fall under geometric deep learning, which embeds geometric inductive biases into DL models for enhanced performance~\autocite{bronstein2021geometricdeeplearning}. By comparing these two models, we aim to investigate whether the neural operator framework, designed to learn mappings between function spaces, offers advantages over the GNN-based approach for this application.

To develop these models, we generate a dataset of 35{,}000 paird input-output samples using finite-element (FE) simulations on synthetically generated LV geometries. Each sample consists of (i) an LV geometry represented as a point cloud on a coarse resolution, (ii) one or two pacing locations, and (iii) a tissue conductivity value as inputs, and the resulting activation time map as the output. The dataset is used to train and test both models, which are subsequently evaluated on real-world LV geometries and finer mesh resolutions.

We show that the trained DL models can rapidly and accurately predict activation patterns across arbitrary LV geometries, tissue conductivities, and pacing sites. We also demonstrate the utility for CRT-motivated pacing analysis by developing a workflow to identify an optimal pacing site on the LV from a given LV geometry and a baseline activation time map. To enhance user accessibility, we also provide an interactive web-based graphical user interface (GUI) built on top of the trained models.

\section{Results}
\subsection{Characteristics of the generated dataset}
\begin{figure}[ht]
\centering
\includegraphics[width=0.99\textwidth]{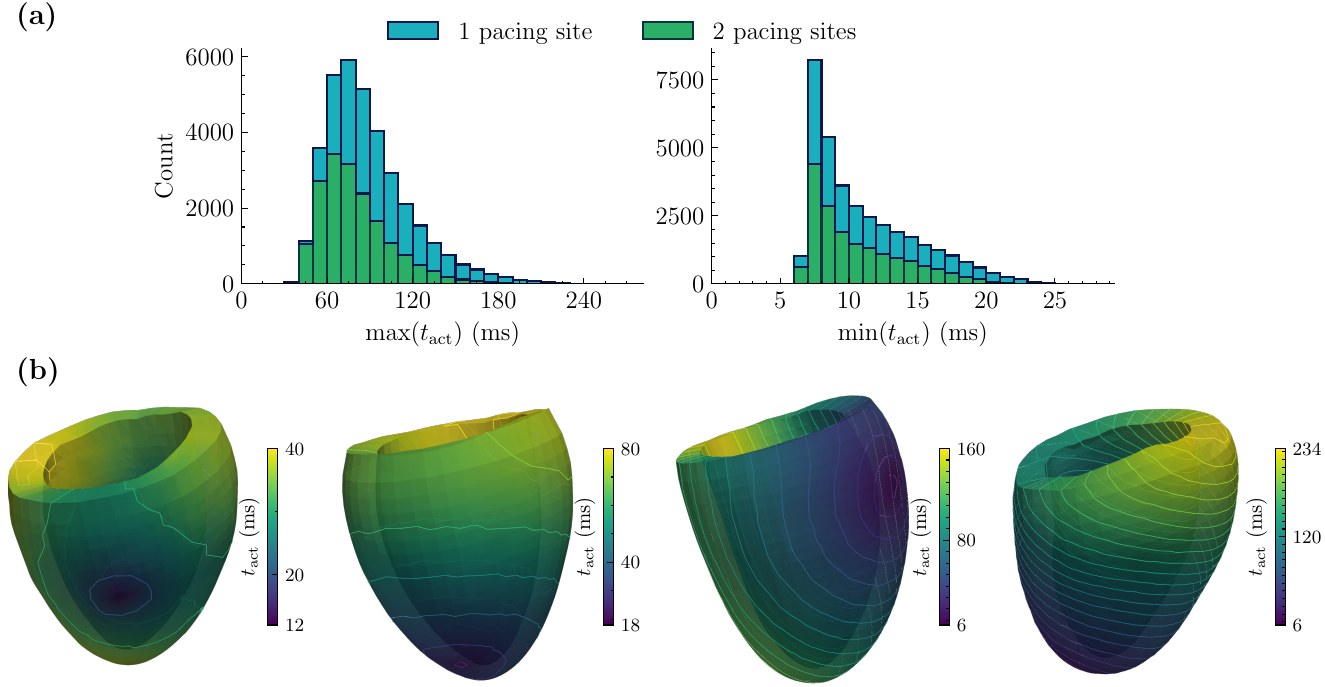}
\caption{FE simulation results. (a) Distributions of minimum and maximum activation times across 35{,}000 FE simulations. (b) Examples of activation patterns. Isocontour lines indicate activation times at integer multiples of 10~ms within the activation range for each case.}
 \label{fig:FEresults}
\end{figure}

The 35{,}000 FE simulations produced a broad range of activation behaviors. \autoref{fig:FEresults}~(a) shows the distributions of minimum and maximum activation times across all cases, while \autoref{fig:FEresults}~(b) presents representative examples of activation patterns. Maximum activation times ranged from 33~ms to 258~ms, with 95\% of cases falling between 48~ms and 164~ms. The earliest activated regions appeared between 6~ms and 25~ms. Half of the simulations corresponded to single pacing sites, representing intrinsic activation, whereas the remaining half included dual pacing sites corresponding to post-CRT scenarios. Each FE simulation required approximately 10~minutes of computation on a 32-core CPU.

\subsection{Performance analysis of GNN and GINO models}
\autoref{tab:training_results} summarizes the training performance of the GNN and GINO models. The GINO model achieved lower training and dev errors than the GNN model while having fewer trainable parameters. Both models were trained on four NVIDIA H200 Tensor Core GPUs (with 140~GB memory each), and their wall-clock training times per epoch were comparable (GNN: 47~s vs. GINO: 55~s).

\begin{table}[ht]
    \caption{Training results for the GNN and GINO models. Columns list the number of trainable parameters, batch size, wall-clock training time per epoch, and errors \(\mathcal{L}\) on the train and dev datasets.}
    \label{tab:training_results}
    \centering
    \begin{tabular}{lcccccc}
    \toprule
    \multirow{2}{*}{model} &
    \multirow{2}{*}{\begin{tabular}[c]{@{}c@{}}\# params \\(million)\end{tabular}} &
    \multirow{2}{*}{\begin{tabular}[c]{@{}c@{}}batch\\size\end{tabular}} &
    \multirow{2}{*}{\begin{tabular}[c]{@{}c@{}}time per \\epoch (s)\end{tabular}} &
    \multicolumn{2}{c}{$\mathcal{L}$} \\
    \cmidrule(lr){5-6}
     & & & & train & dev \\
    \midrule
    GNN & 3.93  & 275 & 47 & 0.0191  & 0.0244 \\
    GINO & 2.78 & 97 & 55 & 0.0100  & 0.0138 \\
    \bottomrule
    \end{tabular}
\end{table}

\begin{figure}[ht]
    \centering
    \includegraphics[width=0.99\textwidth]{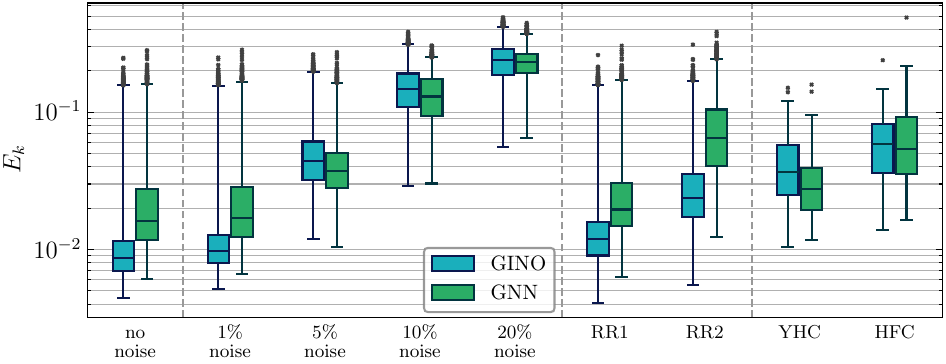}
    \caption{Test set errors for the GNN and GINO models: no noise, Gaussian perturbations of $1\% - 20\%$, RR1 and RR2, and on real-world LV geometries from YHC and the HFC. The synthetic test, RR1, and RR2 sets contain 8,750 cases; YHC and HFC contain 225 and 50 cases, respectively. Whiskers extend to the minimum and 99.5th percentile of observed errors. Remaining outliers are shown individually.}
    \label{fig:test-boxplots}
\end{figure}

\autoref{tab:GNNvsGINO} and \autoref{fig:test-boxplots} summarize test dataset performance, robustness to Gaussian input noise, and sensitivity to input resolution, along with results on real-world LV geometries. Sensitivity to spatial resolution was evaluated by re-meshing the synthetic test set at two finer resolutions (RR1 and RR2), which had approximately 7 and 51 times more nodes than the original test set, respectively. Generalizability to real-world LV geometries was assessed using two clinical cohorts, named young heart cohort (YHC, n=225) and heart-failure cohort (HFC, n=50), derived from a publicly available dataset~\autocite{ChoKim2021}.

\begin{table}[ht]
    \centering
    \caption{Test results of the trained GNN and GINO models. $P_{\le 99.5}$ denotes the largest error in the dataset not exceeding the 99.5th percentile.}
    \label{tab:GNNvsGINO}
    \begin{tabular}{lccccccc}
    \toprule
    \multirow{2}{*}{dataset} &
    \multirow{2}{*}{\# cases} &
    \multicolumn{3}{c}{$\mathcal{L}_{\mathrm{GNN}}$} &
    \multicolumn{3}{c}{$\mathcal{L}_{\mathrm{GINO}}$} \\
    \cmidrule(lr){3-5}
    \cmidrule(lr){6-8}
         & & mean & $P_{\le 99.5}$  & max &
             mean & $P_{\le 99.5}$  & max \\
    \midrule
    test & 8750 & 0.0244 & 0.1603 & 0.2830 & 0.0136 & 0.1574 & 0.2490\\
    RR1  & 8750 & 0.0280 & 0.1723 & 0.3053 & 0.0168 & 0.1585 & 0.2609 \\
    RR2  & 8750 & 0.0779 & 0.2437 & 0.3856 & 0.0318 & 0.1698 & 0.3111 \\
    YHC  & 225  & 0.0324 & 0.1359 & 0.1590 & 0.0444 & 0.1377 & 0.1508 \\
    HFC  & 50   & 0.0779 & 0.4232 & 0.4899 & 0.0636 & 0.2170 & 0.2395 \\
    \bottomrule
    \end{tabular}
\end{table}

Without added noise, the GINO model outperformed the GNN model, achieving nearly half the mean error on the synthetic test set (1.36\% vs. 2.44\%), while their maximum errors were similar (24.9\% vs. 28.3\%). Under Gaussian noise, both models exhibited comparable degradation, with the GNN showing slightly better overall robustness. On the refined resolution sets (RR1 and RR2), both models experienced increases in error relative to the original test set. Because GINO began from a lower baseline error, its absolute errors remained lower across these resolutions (see Supplementary information, \autoref{fig:baseres_VS_RR1RR2}). When evaluated on real LV geometries from YHC and HFC sets, both models showed increased errors. In YHC, the GNN slightly outperformed the GINO model, whereas both models performed similarly on HFC. The aggregated performance of the GNN and GINO models on the real-world datasets combined (YHC + HFC) is 4.07\% and 4.79\%, respectively.

\begin{figure}[ht]
  \centering
  \includegraphics[width=\textwidth]{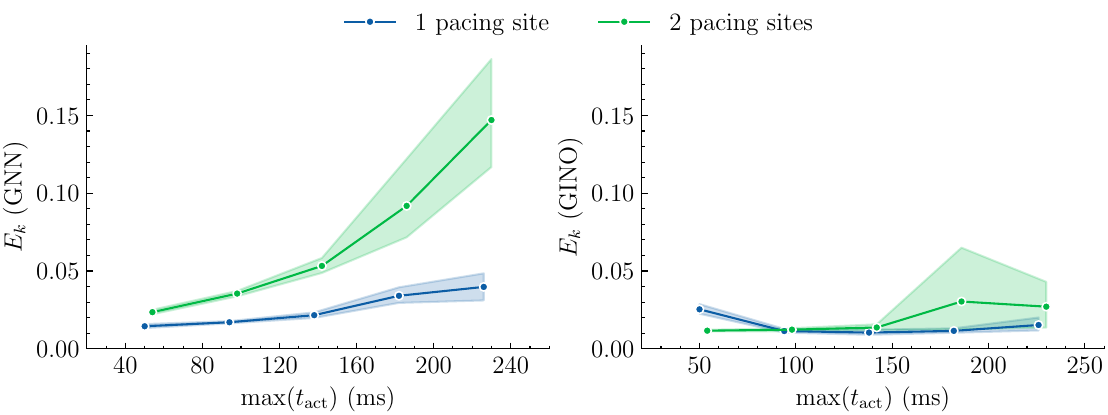}  
  \caption{Variation in error as a function of maximum activation time and number of pacing sites for the noise-free synthetic test set. Left: GNN model. Right: GINO model. Shaded regions indicate 99.5\% confidence intervals of the mean error across subjects.}
  \label{fig:error_t_act}
\end{figure}

\autoref{fig:error_t_act} shows the variation of prediction error across subjects as a function of the maximum activation time and the number of pacing sites in the test set. For the GNN model, errors increase with $\max(t_{\mathrm{act}})$ and are consistently higher for cases with two pacing sites than for single-pacing-site cases. In contrast, the GINO model attains markedly lower errors overall, and its performance is largely comparable between one and two pacing sites, with only a modest increase at very long activation times. All results in this figure are obtained from noise-free test dataset.

\begin{figure}[htp]
\centering
\includegraphics[width=\textwidth]{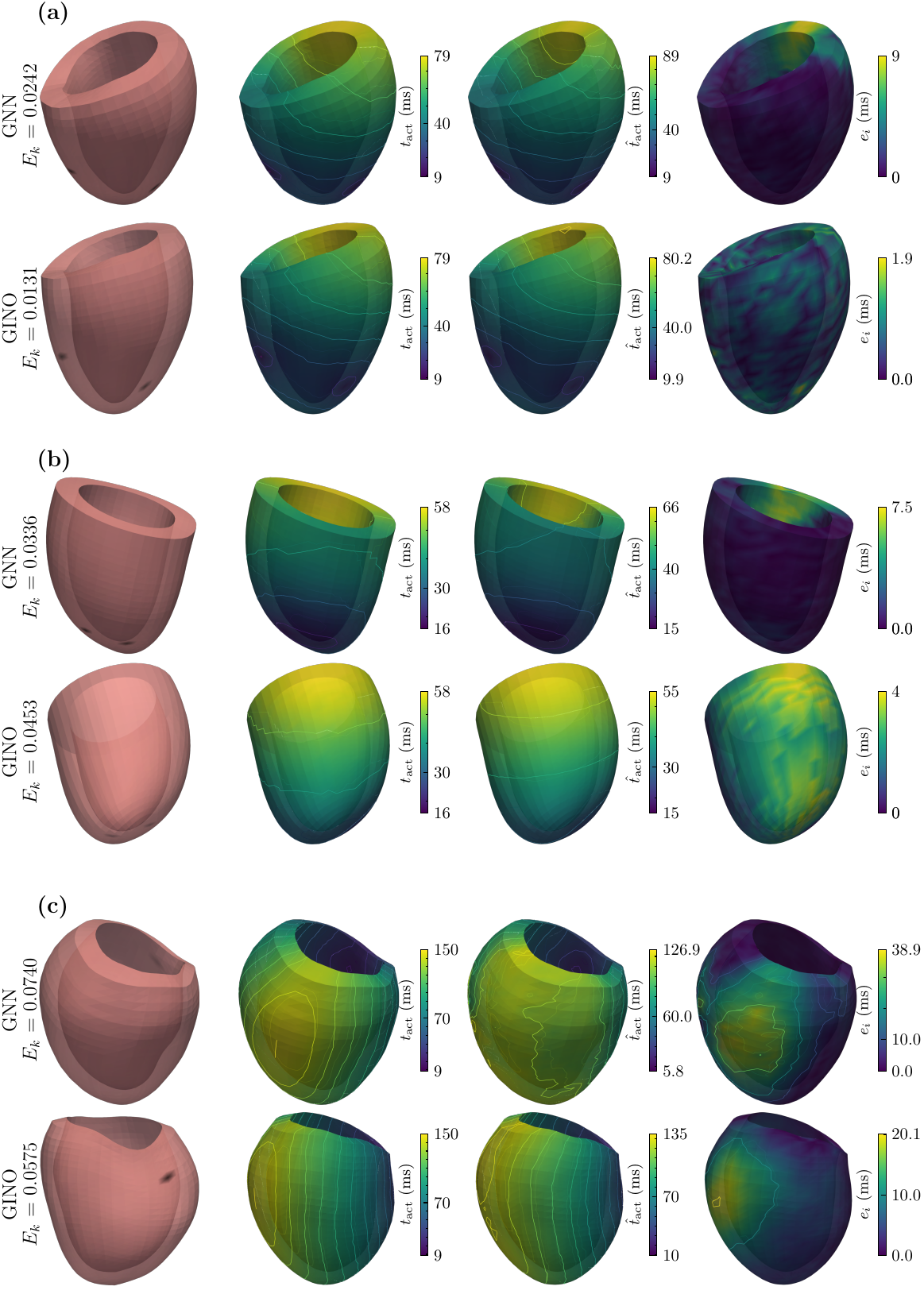}    
\caption{Examples with global error $E_k$ close to the dataset mean. (a) Test dataset, (b) YHC dataset, (c) HFC dataset. In each dataset, a single case with $E_k$ lay near the mean for both models (GNN and GINO) was selected. From left to right, the panels show: the pacing site(s), the activation pattern from the FE simulation ($t_{\mathrm{act}}$), the activation pattern from the DL model prediction ($\hat{t}_{\mathrm{act}}$), and the local error ($e_i$).}
\label{fig:sample_results_mean}
\end{figure}

\begin{figure}[htp]
\centering
\includegraphics[width=\textwidth]{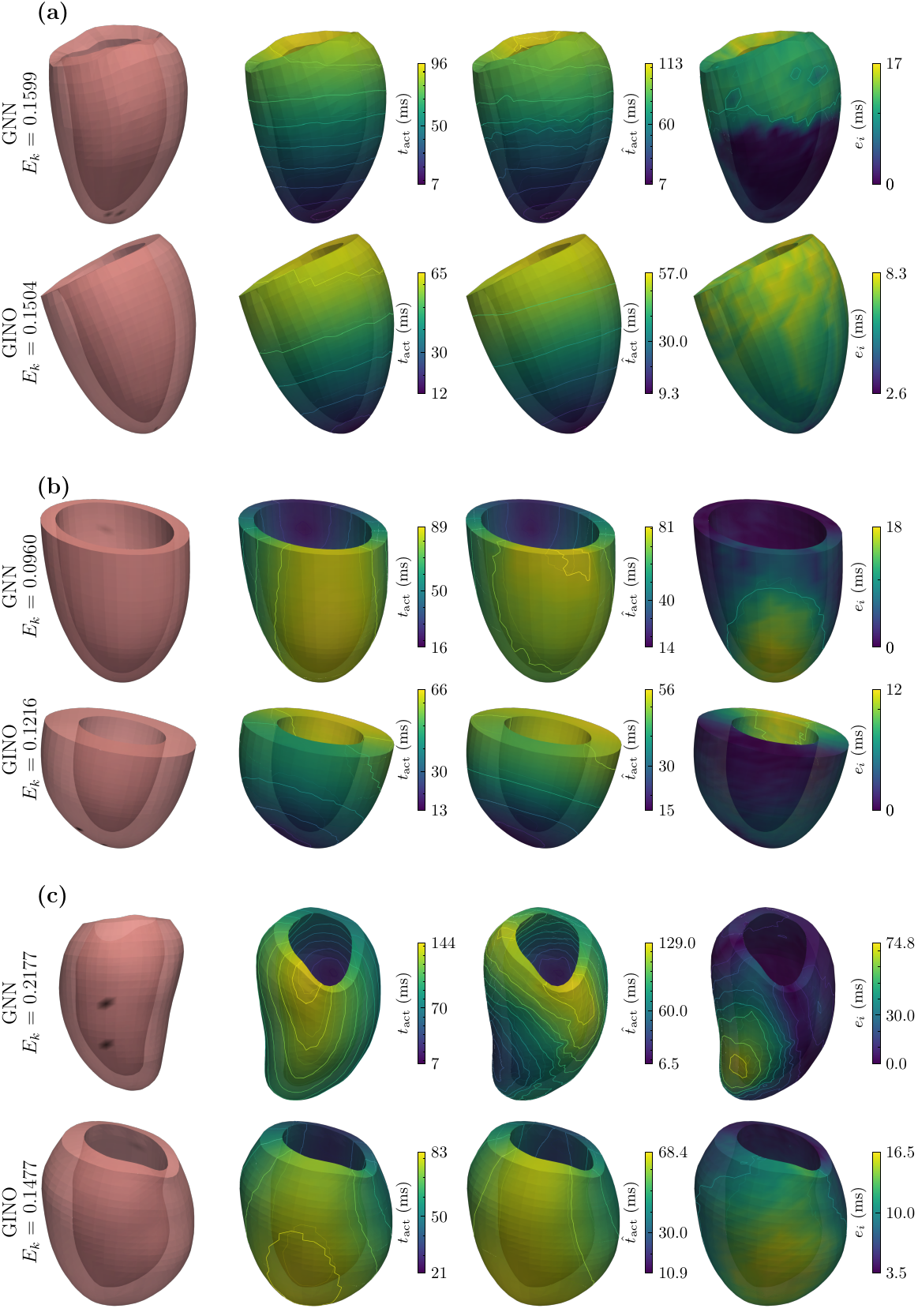}
\caption{High-error examples with global error $E_k$ equal to or just below the 99.5th percentile for each dataset. Refer to \autoref{fig:sample_results_mean} for organization.}
\label{fig:sample_results_99}
\end{figure}

Figs.~\ref{fig:sample_results_mean} and \ref{fig:sample_results_99} compare DL predictions with FE activation maps for cases near the mean error and near the 99.5th percentile across the test, YHC, and HFC datasets. Additional examples, including cases with lower than mean error and with maximum error for each dataset, are provided in Supplementary information, Figs.~\ref{fig:sample_results_below} and \ref{fig:sample_results_max}. Near the mean error level, GINO predictions in overall show closer agreement with FE solutions than GNN predictions. This agreement deteriorates on the HFC dataset, where $E_{k}$ is higher and the geometry deviates more strongly from the training synthetic samples.

In cases with errors near the 99.5th percentile, both models still recovered the overall activation patterns on the test set and on the YHC dataset, although the absolute activation times were inaccurate. On the HFC dataset, however, the GNN model even failed to produce the overall activation patterns, whereas the GINO model continued to reproduce the activation map qualitatively despite exhibiting large absolute errors and a minimum bias of at least 3.5 ms across the domain.

\subsection{Application case: CRT patient simulation}
\begin{figure}[ht]
\centering
\includegraphics[width=\textwidth]{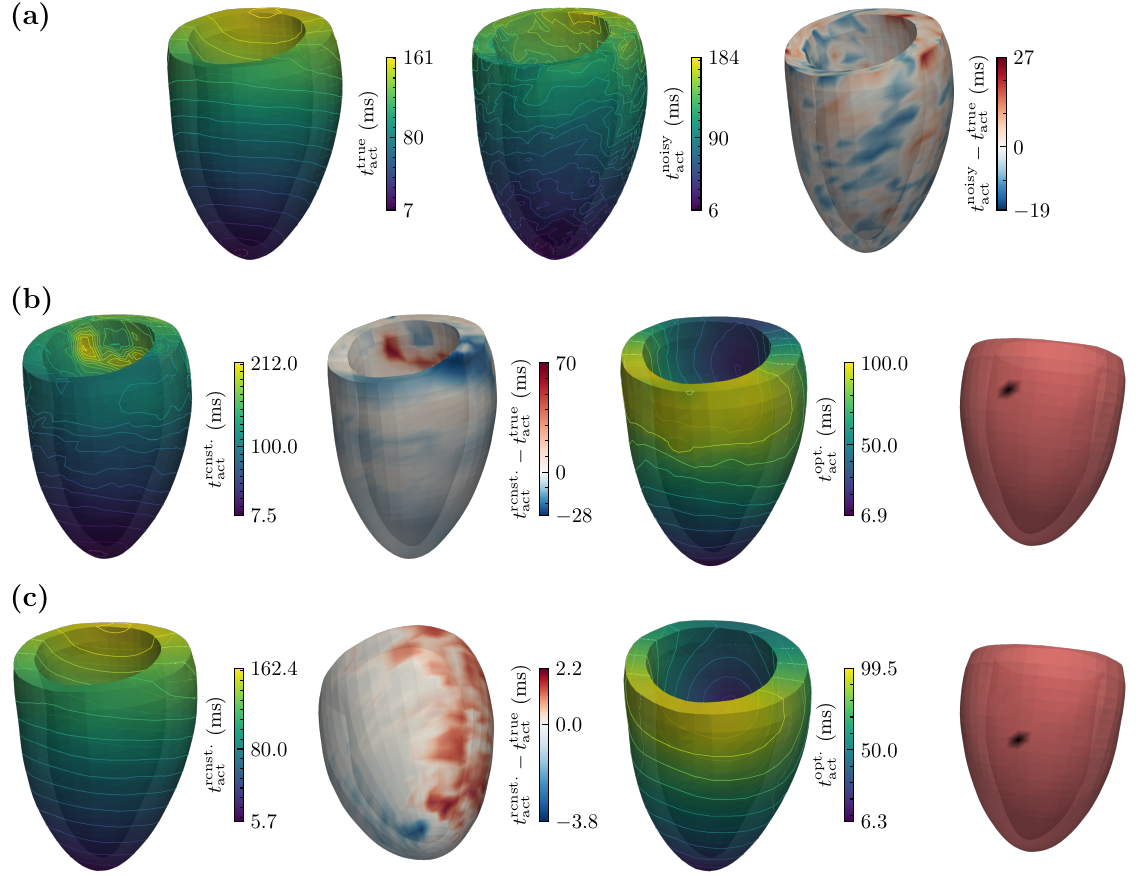}
\caption{Demonstration of the CRT optimization workflow using the trained GNN and GINO models. (a) Ground-truth activation map from the FE simulation (left), noisy activation map used as input to the inverse problem (center), and their difference (right). (b) Results using the GNN model. From left to right: reconstructed activation map after solving the inverse problem, error relative to the ground-truth map, predicted activation map after adding a second pacing site at the optimized location, and the model optimized location for the second pacing site. (c) Results using the GINO model, organized as in (b).}
\label{fig:crt-results}
\end{figure}

We applied the trained GNN and GINO models within our CRT optimization workflow, which identifies an optimal pacing site on the LV that minimizes total activation time for a given LV activation map (\autoref{fig:CRT-workflow}). The corresponding results are shown in \autoref{fig:crt-results} and summarized numerically in \autoref{tab:inv-probCRT}.

\begin{table}[ht]
    \caption{The differences of the estimated parameters in the inverse problem and their difference with the ground truth.}
    \label{tab:inv-probCRT}
    \centering
    \begin{tabular}{lcc}
    \toprule
    \multirow{3}{*}{model} &
    \multirow{3}{*}{\begin{tabular}[c]{@{}c@{}}$d_{\mathrm{iso}}^{\ast}$\\(Error \%)\end{tabular}} &
    \multirow{3}{*}{\begin{tabular}[c]{@{}c@{}}$\left\lVert p_{\text{pacing}_1}^{\ast} - p_{\text{pacing}_1}^{\mathrm{true}} \right\rVert$\\(mm)\end{tabular}} \\
     & & \\
     & & \\
    \midrule
    GNN  & 6.48\%  & 2.1 \\
    GINO & 1.45\%  & 1.6 \\
    \bottomrule
    \end{tabular}
\end{table}

The synthetic test case selected to illustrate the workflow featured a single intrinsic pacing site and a prolonged maximum activation time of 161~ms, reflecting pathological conditions associated with left bundle branch block (LBBB)~\autocite{Surkova2017}. After adding the noise, the maximum activation time increased to 184~ms. \autoref{fig:crt-results}~(a) displays the ground-truth activation map from the FE simulation, the noisy activation map used as input to the inverse problem in the workflow, and their difference. In the forward FE simulation, the endocardial and epicardial fiber angles ($\alpha_{\mathrm{endo}}$, $\alpha_{\mathrm{epi}}$) for this case were $45.3^\circ$ and $-69.9^\circ$, respectively, but in the CRT optimization workflow they were assumed to be $60^\circ$ and $-60^\circ$.

In a clinical setting, the noisy activation map would be replaced by a patient-specific measured map. As shown in \autoref{tab:inv-probCRT}, both models recovered the intrinsic pacing site $p_{\text{pacing}_1}^{\ast}$ and the isotropic conductivity $d_{\mathrm{iso}}^{\ast}$ with small errors despite the added noise. The GINO model showed more robust recovery, and therefore the reconstructed activation map by this model closely matched the ground truth, whereas the GNN model exhibited larger local discrepancies. Post-CRT predictions of both models are close to each other despite their optimal $p_{\text{pacing}_2}^{\ast}$ having a 16.4~mm distance from each other. The whole end-to-end CRT optimization workflow took $\approx$ 25~seconds using the GINO model and $\approx$ 1.5~minutes using the GNN model on one NVIDIA V100 GPU with 32~GB memory, including 5{,}000 forward model evaluations.

\section{Discussion}
The key finding of this study is that the two DL models trained on synthetic LV geometries can predict activation time maps with reasonable accuracy when evaluated on unseen real-world LV geometries. Both architectures also demonstrate potential for individualized CRT optimization, recovering subject-specific parameters from noisy activation time maps with low error.

On the synthetic test set, the GINO model achieves substantially lower error than the GNN (0.0136 vs 0.0244), indicating a clear advantage when training and testing distributions match. However, on the real-world LV geometries from the YHC and HFC datasets, the two models show similar errors (GNN: 0.0407, GINO: 0.0479). This suggests that, in the present setting, performance on clinical anatomies is primarily limited by the mismatch between the synthetic training distribution and the real patient geometries, rather than by model capacity. In a pilot hyperparameter study on an earlier version of the synthetic dataset (see Supplementary information, \autoref{tab:param_opt}), increasing GINO capacity by varying the number of Fourier modes and using a higher rank (0.4 vs. 0.01 in the main study) reduced error, whereas increasing the GNN embedding width did not consistently improve performance over the tested range. The absence of a corresponding gain on real-world LVs for GINO, despite lower error on synthetic data, indicates that further improvements are more likely to come from reducing the synthetic-to-real gap than from adopting a more expressive DL surrogate.

In the current implementation, the GNN's forward pass is more time-consuming than the GINO's, due to repeated downsampling and neighborhood-search operations in each layer. However, the GNN has a lower memory footprint, enabling substantially larger batch sizes. The GINO model, while faster per inference call, requires more memory because of the Fourier transform operations in float32, which restricts its batch size. Overall, the training epoch times were comparable (GNN: 47~s with batch size 275; GINO: 55~s with batch size 97). This trade-off between per-call speed and memory becomes most relevant when the models are embedded in an optimization loop requiring thousands of forward evaluations, such as the CRT planning workflow.

At the subject level (Figs.~\ref{fig:sample_results_mean} and \ref{fig:sample_results_99}), GINO predictions are noticeably smoother, with fewer local sharp edges, whereas GNN predictions contain more abrupt discontinuities. This likely reflects the spectral filtering in GINO, which suppresses high-frequency components, an effect amplified by the low-rank tensor factorization of the Fourier weights (0.01) used to reduce the number of trainable parameters.

In this study, the graph neural operator (GNO) encoder of the GINO model was learned during training. An alternative approach is to use a predefined function like a fixed signed-distance-function encoder and apply the GNO only to the decoder, particularly when the latent grid size ($28^3$ here) is comparable to or exceeds the input mesh resolution (4{,}804 nodes), in which case a learned encoder may become redundant~\autocite{li2023geometryinformed}. In this study, we opted for the full learned encoder-decoder. This choice is not expected to substantially affect model accuracy.

Compared with the neural operator framework proposed by Yin \textit{et al.}~\autocite{DIMON2024}, which was trained on monodomain FE simulations and reported an error of 0.045, both of our models perform comparably (GNN: 0.0407, GINO: 0.048) despite never seeing real-world shapes during training. Their formulation was designed for a more constrained pacing configuration (seven discrete stimulation sites, no dual-pacing) and assumes constant conductivity across subjects. We note that the model of Yin \textit{et al.} is substantially more lightweight to train (single GPU, 2 GB memory vs. our four NVIDIA H200 GPUs with 140 GB each), although inference for all models is effectively instantaneous. A further distinction is that our models operate directly on point-cloud LV geometries, while Yin \textit{et al.} employ PCA-based geometric encoding, which may limit representation of fine local anatomical features.

\begin{figure}[ht]
    \centering
    \includegraphics[width=0.98\textwidth]{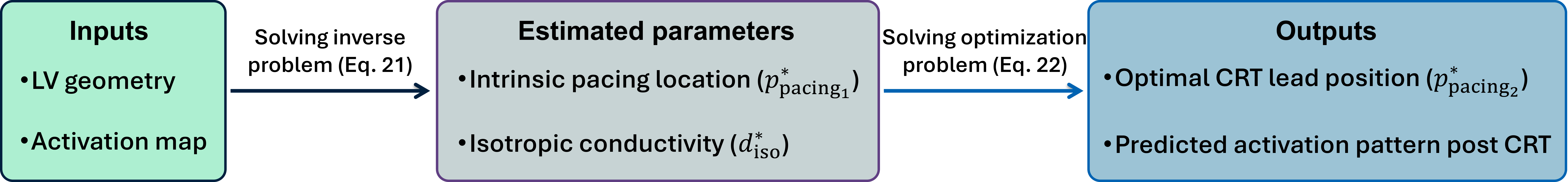}
    \caption{CRT optimization workflow.}
    \label{fig:CRT-workflow}
\end{figure}

The computational efficiency of the trained DL models, which predict activation maps in just a few milliseconds (cf. to $\sim$ 10~min in FE simulations), makes them suitable for inverse and optimization problems requiring large numbers of forward evaluations. To demonstrate this potential utility, we developed a workflow that identifies the optimal lead position with the LV geometry and corresponding activation time map as inputs (\autoref{fig:CRT-workflow}). We note that the LV geometry and the activation time map can be obtained from electroanatomical mapping~\autocite{Narayan2024} in clinical practice. Both the GNN and GINO models accurately recovered the intrinsic pacing site $p_{\text{pacing}_1}^{\ast}$ and tissue conductivity $d_{\mathrm{iso}}^{\ast}$ from noisy activation time maps (\autoref{tab:inv-probCRT}). The GINO model was additionally able to reconstruct the ground-truth activation map from the noisy input with negligible error (\autoref{fig:crt-results}).

Fiber angle variations had negligible effects on the predicted activation maps. Nearly identical activation patterns were obtained even when the fiber orientations differed between the forward simulation ($\alpha_{\mathrm{endo}} = 45.3^\circ$, $\alpha_{\mathrm{epi}} = -69.9^\circ$) and the inverse problem ($\alpha_{\mathrm{endo}} = 60^\circ$, $\alpha_{\mathrm{epi}} = -60^\circ$), consistent with prior work~\autocite{RuppMyocardialFiber2021, Pravdin2014} noting minimal impact within $\pm 20^\circ$ of variation.

Post-CRT predictions were similar between models despite their optimal pacing sites $p_{\text{pacing}_2}^{\ast}$ being 16.4~mm apart. The models are relatively insensitive to pacing-site perturbations within a certain spatial range. Based on sweeps over pacing locations, changes of roughly $>$~20~mm in geodesic distance were required to produce $>$~5~ms differences in the maximum activation time. Both models typically select locations near the latest activated region at baseline, which is expected given the assumption of homogeneous conductivity. The complete CRT workflow finishes in 25 seconds using the GINO model, even though it solves two optimization problems (Eqs.~\ref{eqn:inv_prob}-\ref{eqn:crt_opt}) with ~5,000 evaluations.

To our knowledge, this overall level of efficiency has not yet been achieved by state-of-the-art physics-based solvers. Although our DL models currently predict activation time maps only, the same architectures could be extended to output additional fields derived from the monodomain simulations with minimal additional cost at inference. For comparison, eikonal-based solvers, which are substantially more efficient than monodomain simulations and compute only activation times, still require runtimes of around 7~s for meshes with $\sim 2$ million degrees of freedom~\autocite{NEIC2017191} and $>1$~s for meshes with 266k elements~\autocite{Ganellari2020}. In contrast, the inference time of the DL surrogates is decoupled from the FE mesh resolution used to generate the training data and remains in the millisecond range once the models are trained. While prediction errors increase when evaluated on finer meshes, the GINO architecture offers additional capacity (e.g., higher tensor ranks) that can be used to further reduce this error without substantially increasing training complexity or changing the inference time.

To increase the accessibility of the trained models, we also developed a web-based interactive GUI (\url{https://dcsim.egr.msu.edu/}) allowing users to upload patient-specific LV geometries, evaluate pacing scenarios interactively in real time, visualize activation maps, and run the CRT optimization workflow. This demonstrates the model's potential to be further extended toward a practical clinical decision-support tool. This developed workflow sets the foundation for \textit{in-silico} optimization of CRT, and continuous updates to this GUI will be provided as newer, more expressive backend DL models are developed.

The main limitation is that the models were trained exclusively on synthetic LV geometries and evaluated on a limited set of real-world LVs without clinical outcome data. As a result, their generalizability to broader patient populations and their impact on prospective CRT decision-making remain to be established. Beyond the data limitation, we highlight three modeling limitations, which point to areas for future improvement. First, the DL models focus only on cardiac electrophysiology and do not consider mechanical deformation. As a result, the CRT workflow depends on the LV electroanatomical activation time map as input, which is difficult to obtain and not routinely collected clinically. Reconstruction of 3D geometry from these data is also challenging. Future work will incorporate electromechanics to enable the use of strain or displacement data. Such models may be trained on synthetic datasets from fully coupled simulations or via physics-informed strategies that reduce the need for precomputed data~\autocite{RAISSI2019}. Incorporating differential equation constraints into neural operator architectures such as GINO can be technically challenging~\autocite{PINOforPDE,Goswami2023}, though recent developments are promising~\autocite{lin2025mollified}. Second, the present study considers only LV geometry. For clinical relevance, full biventricular (BiV) geometry should be included. In addition, our pacing-site optimization is performed over the entire LV surface, whereas in practice feasible pacing locations are substantially constrained. This can be addressed by incorporating an anatomical mask reflecting clinically accessible lead positions. We also did not include an explicit model of the fast-conduction Purkinje fiber network. In LBBB patients, however, the LV is largely activated via slow cell-to-cell ventricular conduction rather than via the Purkinje system~\autocite{Vajapey2023}. Nevertheless, future models including a Purkinje representation may better capture its effects on activation patterns~\autocite{Gillette2021}. Third, tissue conductivity was assumed homogeneous, whereas scar and ischemic regions introduce spatial heterogeneity that can affect the activation patterns and optimal lead placement. We also assumed that the activation time map associated with LBBB can be reproduced from a single pacing site, which may not be true. Future work will incorporate additional pacing locations and local variation in tissue conductivity to better capture these clinically relevant effects.

In summary, we compared two physics-based DL models, namely Graph-UNet based GNN and GINO, and showed that both models can predict activation patterns across diverse LV geometries and pacing configurations. We further demonstrated the potential clinical utility of this study for \textit{in-silico} optimization of the pacing site in CRT planning and its accessibility through an interactive web-based GUI that enables users to visualize and evaluate different pacing scenarios in real time. This study sets the foundation for future electromechanical modeling and extension to biventricular geometries.

\section{Methods}
\begin{figure}[ht]
    \centering
    \includegraphics[width=.9\textwidth]{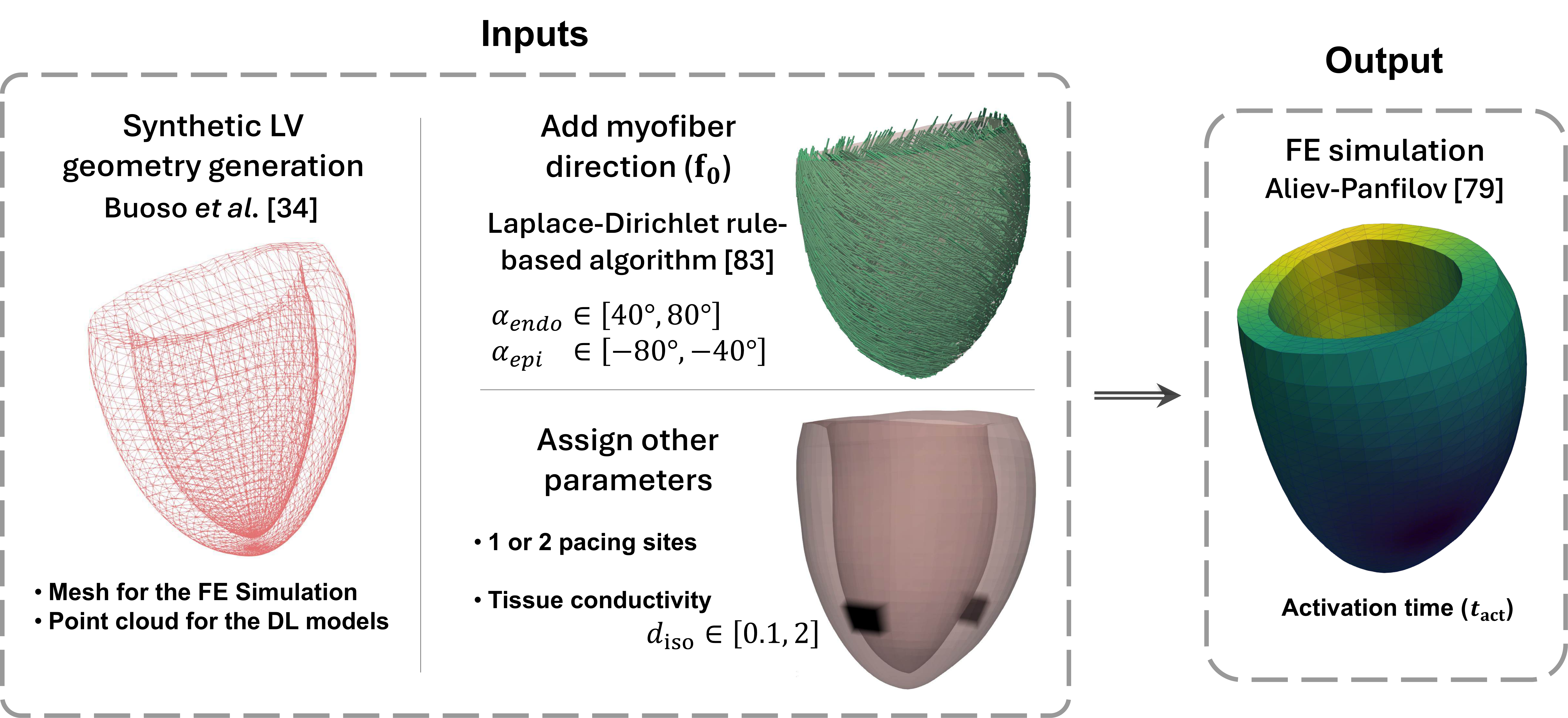}
    \caption{Generation of input-output dataset for training the DL models. A synthetic LV geometry is generated using the method described by Buoso \textit{et al.}~\autocite{BUOSO2021102066}. The mesh is used for the FE simulation, whereas only the corresponding point cloud is retained for the DL models. LV myofiber direction is computed using a Laplace-Dirichlet rule-based algorithm~\autocite{Bayer2012}, where endocardial and epicardial fiber angles ($\alpha_{\mathrm{endo}}$, $\alpha_{\mathrm{epi}}$) are randomly assigned within the ranges of $[40^\circ, 80^\circ]$ and $[-80^\circ, -40^\circ]$, respectively. One or two pacing sites and a random subject-specific tissue electrical conductivity value are then selected. The activation pattern is computed by solving the Aliev-Panfilov model~\autocite{Aliev1996} using the FE method. A total of 35,000 simulations were conducted to generate the dataset used for training the DL models.}
    \label{fig:prob-statement}
\end{figure}

We generated a dataset of 35{,}000 paired input-output samples using a multi-stage pipeline involving synthetic LV geometry generation, fiber assignment, pacing selection, conductivity sampling, and FE electrophysiology simulation. An overview of the full data generation workflow is shown in \autoref{fig:prob-statement}. Detailed methodology is provided in the subsections below.

\subsection{LV geometries}
\begin{figure}[ht]
    \centering
    \includegraphics[width=\textwidth]{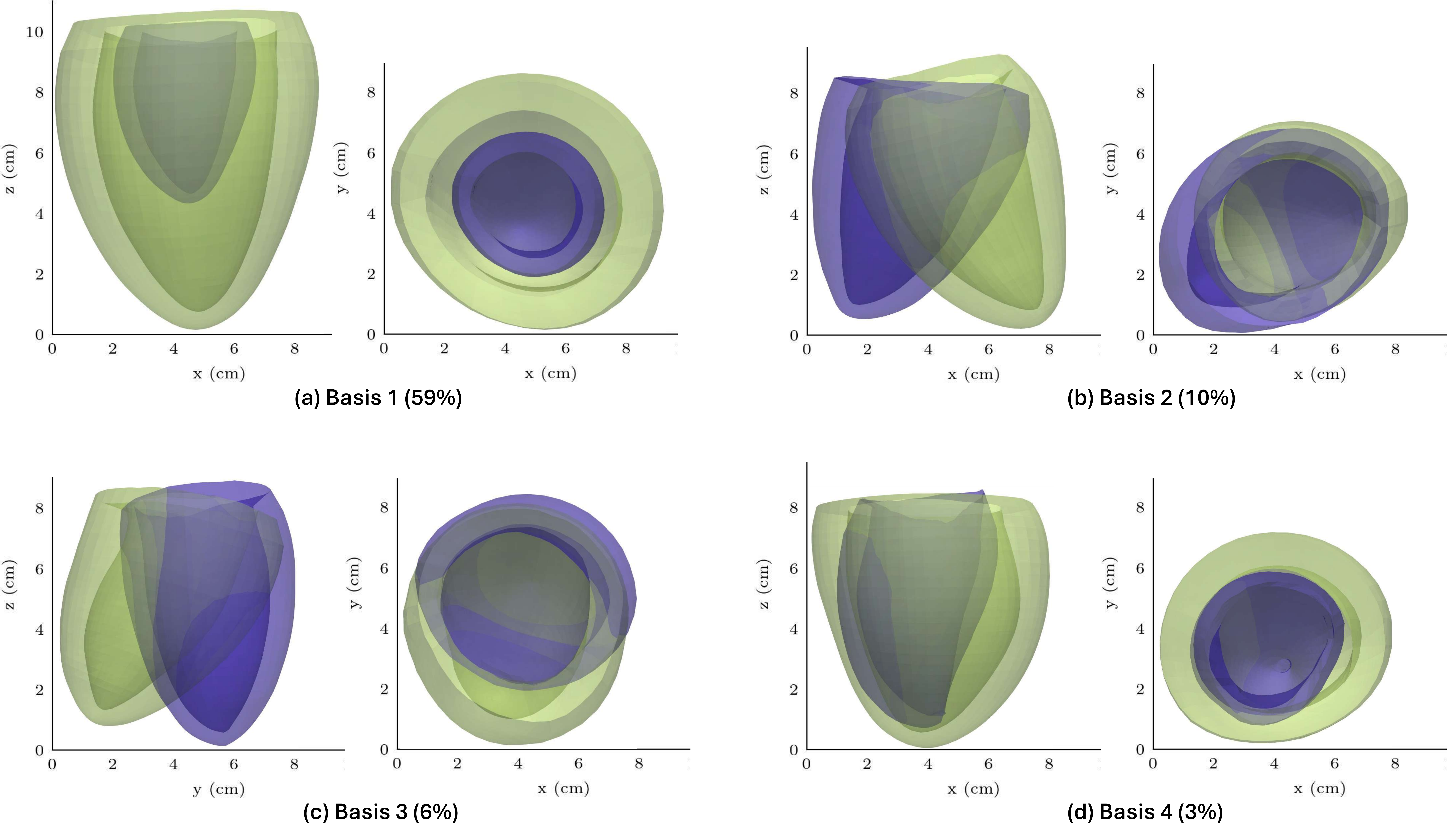}
    \caption{Illustration of the first four POD bases. Each panel illustrates the impact of adding the respective basis to the mean shape, using weights corresponding to the maximum (green) and minimum (blue) amplitudes identified by the POD analysis. Percentages in brackets indicate the variance represented by each base.}
    \label{fig:POD-bases}
\end{figure}

Synthetic geometries were generated following the methodology proposed by Buoso \textit{et al.}~\autocite{BUOSO2021102066}, which employs proper orthogonal decomposition (POD) to construct a statistical shape model from 75 segmented LV geometries spanning healthy and pathological cases (ages 5-80 years). The POD model maps a reference LV mesh (21{,}240 tetrahedral elements, 4{,}804 nodes) to capture geometric variations of the LV within the dataset. \autoref{fig:POD-bases} shows the variations associated with the first four POD bases. Synthetic geometries were generated by random sampling of the POD coefficients within the ranges defined by the original study. In total, 35{,}000 geometries were created for training, development (hyperparameter optimization), and testing of the DL models. Further details on geometry generation and quality control are provided in Supplementary information.

For co-registration of geometries, Buoso \textit{et al.} provided a shared anatomical frame with a physiological parametrization (transmural $x_t \in [0, 1]$, circumferential $x_c \in [-\pi, \pi]$, and longitudinal $x_l \in [0, 1]$ coordinates) and a common rigid alignment following previous studies~\autocite{TOUSSAINT20131243, PAUN2017270}. The transmural coordinate was defined by assigning 0 and 1 on the endocardium and epicardium, respectively. An eikonal problem was solved to compute the distance from the apex~\autocite{Fares2002}, and the resulting field was normalized circumferentially. Each mesh was translated so that the basal endocardial centroid lay at the origin, rotated so that the basal-plane normal aligned with the $+z$ (apex-base) direction, and then rotated about the $z$-axis so that the inferior intersection between left and right ventricles intersected the $-x$ axis.

Because the baseline mesh resolution was coarse for electrophysiological FE simulations, two uniform refinements were applied by subdividing each tetrahedral element into eight smaller ones, yielding meshes with 169{,}920 elements (33{,}327 nodes) after the first refinement and 1{,}359{,}360 elements (246{,}493 nodes) after the second refinement. All simulations were performed on the second refinement (see Supplementary information, Mesh independence and time step sensitivity studies). The resulting activation maps were downsampled to the coarse mesh (4{,}804 nodes) for memory efficiency and training. Therefore, the DL models only saw the coarse resolution during training. Post-training, activation maps in the test dataset were interpolated onto both refined meshes, referred to as refined resolution 1 (RR1, 33{,}327 nodes) and refined resolution 2 (RR2, 246{,}493 nodes), to assess discretization invariance of the DL models.

To evaluate the trained models on real-world LV anatomies, we used LV geometries from a publicly available dataset~\autocite{ChoKim2021} comprising 50 patients with heart failure selected for CRT and 225 young individuals (age 26.4 $\pm$ 2.1 years). These two cohorts are referred to as the heart failure cohort (HFC) and the young heart cohort (YHC), respectively. The dataset provides surface meshes. Using the same preprocessing as Buoso \textit{et al.}, we co-registered these geometries to the same coordinate system and performed tetrahedralization to obtain volumetric meshes with the same baseline resolution (4{,}804 nodes).

\subsection{Governing equations of cardiac electrophysiology}
Cardiac electrical activity and its propagation in the LV was described using the Aliev-Panfilov model~\autocite{Aliev1996}. This model employs two dimensionless state variables: the cardiac action potential $\phi$ and a recovery variable $\eta$, governed by
\begin{subequations} \label{eqn:aliev_panfilov}
    \begin{align}
        \dfrac{\partial \phi}{\partial t} &= \mathbf{\nabla} \cdot (\mathbf{D} \mathbf{\nabla} \phi) + f_{\phi}(\phi, \eta) + I_s, \\
        \dfrac{\partial \eta}{\partial t} &= f_{\eta}(\phi, \eta),
    \end{align}
\end{subequations}
\noindent on the domain $\Omega$ and its surface $\Gamma$ subject to
\begin{subequations} \label{eqn:bcs_ics}
    \begin{align}
        \dfrac{\partial \phi}{\partial \mathbf{n}} = 0, &\qquad \text{on } \Gamma,\\
        \phi \text{ and } \eta = 0, &\qquad \text{at } t=0,
    \end{align}
\end{subequations}
\noindent where $t$ denotes dimensionless time, set to zero at electrical rest (end of the diastolic interval). The nondimensionalization of $\phi$ and $t$ is
\begin{align}
    \phi = \dfrac{V_m + 80}{100}, \qquad t = \dfrac{t_{\mathrm{phys}}}{12.9}
\end{align}
where $V_m$ is the transmembrane potential in mV, and $t_{\mathrm{phys}}$ is time in ms. This maps $V_m\in[-80,20]$ mV to $\phi\in[0,1]$ and matches the model action potential duration of $25.58$ to $\sim\!330$ ms~\autocite{Nash2004}.

The anisotropic electrical conductivity tensor is defined as
\begin{align} \label{eqn:elec-cond}
    \mathbf{D} = d_{\mathrm{iso}} \mathbf{I} + d_{\mathrm{ani}} \mathbf{f}_{0} \otimes \mathbf{f}_{0},
\end{align}
with $d_{\mathrm{iso}}$ and $d_{\mathrm{ani}}$ denoting the isotropic and anisotropic conductivities, respectively, and $\mathbf{f}_{0}$ denoting the myofiber direction vector. To reflect faster conduction along fibers, we set $d_{\mathrm{ani}} = 4d_{\mathrm{iso}}$, in line with literature values~\autocite{LORANGE1993245, PLONSEY1984557}. In our simulations, $d_{\mathrm{iso}}$ was treated as a subject-specific parameter, and its value was randomly selected from the interval $[0.1, 2]$ to capture a wide range of outputs. The beta distribution with shape parameters $\alpha = 1.2$ and $\beta = 3$ was used to favor lower $d_{\mathrm{iso}}$ values, which are more representative of heart failure patients. The myofiber direction $\mathbf{f}_{0}$ was prescribed using a Laplace-Dirichlet rule-based algorithm~\autocite{Bayer2012}, with a linear transmural variation of the helix angle from the endocardium to the epicardium. For each case, the endocardial and epicardial fiber angles ($\alpha_{\mathrm{endo}}$, $\alpha_{\mathrm{epi}}$) were independently sampled from uniform distributions within $[40^\circ, 80^\circ]$ and $[-80^\circ, -40^\circ]$, respectively.

Local excitation was initiated by randomly selecting one or two pacing sites (radius of 1 mm). A constant dimensionless electrical stimulus $I_s$ with a magnitude of 25 was applied at these locations between $t_{\mathrm{phys}} = 4$ ms and $t_{\mathrm{phys}} = 24$ ms. Single pacing-site simulations represented intrinsic activation associated with LBBB, whereas dual pacing-site simulations modeled post-CRT scenarios, where the second pacing site represents the placed lead. The excitation properties of cardiac tissue are defined by
\begin{subequations} \label{eqn:f_phi_r}
    \begin{align}
        f_{\phi}(\phi, \eta) &= -c \phi (\phi - \alpha) (\phi - 1) - \eta \phi, \\
        f_{\eta}(\phi, \eta) &= \left(\gamma + \dfrac{\mu_1 \eta}{\mu_2 + \phi}\right) \left(-\eta - c \phi (\phi - b -1)\right),
    \end{align}
\end{subequations}
with the parameters associated with cellular-level dynamics set to $c = 8$, $\alpha = 0.01$, $\gamma = 0.002$, $\mu_1 = 0.2$, $\mu_2 = 0.3$, and $b = 0.15$, following~\autocite{Aliev1996}. Patient variability of these values was neglected. The activation time $t_{\mathrm{act}}$ at each spatial location $\mathbf{x} = (x, y, z)$ is computed as
\begin{align} \label{eqn:t_act}
    t_{\mathrm{act}}(\mathbf{x}) = \inf\big\{t_{\mathrm{phys}}(\mathbf{x}) | \phi(\mathbf{x}, t) \ge 0.9\big\}.
\end{align}

\noindent The governing equations \autoref{eqn:aliev_panfilov} were solved using the FE method~\autocite{FAN2022105050, Arumugam2019} implemented in \emph{FEniCS legacy}~\autocite{AlnaesEtal2015, LoggEtal2012, LoggWells2010}. The weak form of the discretized problem is expressed as 
\begin{align} \label{eqn:ap-weakform}
    F &= \int_{\Omega} \dfrac{\phi - \phi_n}{k} \phi_{\mathrm{test}} d\Omega - \int_{\Omega} \mathbf{\nabla} \cdot (\mathbf{D} \mathbf{\nabla}\phi) \phi_{\mathrm{test}} d\Omega - \int_{\Omega} f_{\phi}(\phi, \eta) \phi_{\mathrm{test}} d\Omega - \int_{\Gamma} I_{s} \phi_{\mathrm{test}} dS,
\end{align}
which is satisfied for all test functions $\phi_{\mathrm{test}} \in H^{1}(\Omega)$. Time integration for $\phi$ used an implicit backward Euler scheme with a dimensionless time step $k \approx 0.0775$ (1~ms), while the recovery variable $\eta$ was advanced explicitly in a staggered scheme by a forward-Euler step.

\subsection{Geometric DL surrogate modeling}
\paragraph{Input and output features.} The input features, denoted by $\msfsl{a}$, include spatial coordinates $(x, y, z)$ of a point cloud in the LV wall, a Boolean variable $p_{\text{pacing}}$ specifying whether a point lies within a pacing site, the isotropic tissue electrical conductivity $d_{\mathrm{iso}}$, and the myofiber direction vector $\mathbf{f}_{0} = (\mathrm{f}_{0_{x}}, \mathrm{f}_{0_{y}}, \mathrm{f}_{0_{z}})$. The output, denoted by $\msfsl{f}$, is the activation time $t_{\mathrm{act}}$ corresponding to the point cloud. Specifically, for each point, we have
\begin{align} \label{eqn:features}
    \msfsl{a} = (x, y, z, p_{\text{pacing}}, d_{\mathrm{iso}}, \mathrm{f}_{0_{x}}, \mathrm{f}_{0_{y}}, \mathrm{f}_{0_{z}}), \qquad \text{ and } \qquad \msfsl{f} = t_{\mathrm{act}}.
\end{align}

In this study, $d_{\mathrm{iso}}$ remains homogeneous within each subject. The spatial coordinates $(x, y, z)$, conductivity ($d_{\mathrm{iso}}$), and activation times ($t_{\mathrm{act}}$) are scaled to Gaussian distributions with zero mean and unit variance.

\paragraph{GNN model.} The GNN architecture employed in this study is based on Graph-UNet~\autocite{gao2019graphunets}, an extension of the widely used UNet architecture that was originally proposed for image segmentation ~\autocite{UNet2015}. UNet effectively captures global context through downsampling while preserving local information through skip connections during upsampling. Graph-UNet adapts this approach to graph-structured data, enabling efficient learning on non-Euclidean domains.

We adopted the Graph-UNet implementation by Bonnet \textit{et al.}~\autocite{bonnet2022an}, developed for solving steady-state incompressible Navier-Stokes equations (\autoref{fig:graphunet-Arch}). Network hyperparameters were chosen similarly, with specific adjustments tailored to our cardiac activation prediction task.

\begin{figure}[ht]
    \centering
    \includegraphics[width=0.98\textwidth]{fig9.pdf}
    \caption{Graph-UNet architecture.}
    \label{fig:graphunet-Arch}
\end{figure}

Initially, node features are mapped to an embedding dimension of 64 through a four-layer multilayer perceptron (MLP) with 256 neurons per layer. Subsequently, node embeddings are updated using SageConv (GraphSAGE) layers~\autocite{graphsage2017}, which aggregate information from neighboring nodes as follows
\begin{align} \label{eqn:SageConv}
    \mathbf{z}_{i}^{(\ell)} = \mathbf{W}_1 \mathbf{z}_{i}^{(\ell - 1)} + \mathbf{W}_2 \left(\dfrac{1}{\left|\mathcal{N}(i)\right|} \sum_{j \in \mathcal{N}(i)} \mathbf{z}_{j}^{(\ell-1)}\right),
\end{align}
where $\mathbf{z}_{i}^{(\ell - 1)}$ and $\mathbf{z}_{i}^{(\ell)}$ are the input and output embeddings of node $i$ at layer $\ell$, respectively, $\mathcal{N}(i)$ denotes the set of neighboring nodes, and $\mathbf{W}_1$ and $\mathbf{W}_2$ are learnable weight matrices.

The initial neighborhood radius was set to 5 mm, close to the mean edge length of the underlying mesh, resulting in approximately 174{,}000 connectivity sizes per case on average. In the downsampling path, after each SageConv layer, a fraction of points is retained randomly (specifically, $\tfrac34$, $\tfrac34$, $\tfrac23$, and $\tfrac23$), and connectivity is recalculated with updated neighborhood radius $r$ of 7.5, 12.5, 17.5, and 25 mm, respectively. During training, a maximum of 32 neighbors per node is considered to reduce computational cost, while during inference, this limit increases to 128 neighbors to capture richer local information and improve prediction accuracy.

Each SageConv layer in the downsampling path doubles the channel dimension, which is subsequently halved in the upsampling path. Skip connections ensure detailed local information is preserved during the reconstruction phase. Each SageConv layer is followed by batch normalization and a GeLU activation function. A final decoder MLP maps the node embeddings back to the scalar activation times.

\begin{figure}[ht]
    \centering
    \includegraphics[width=0.98\textwidth]{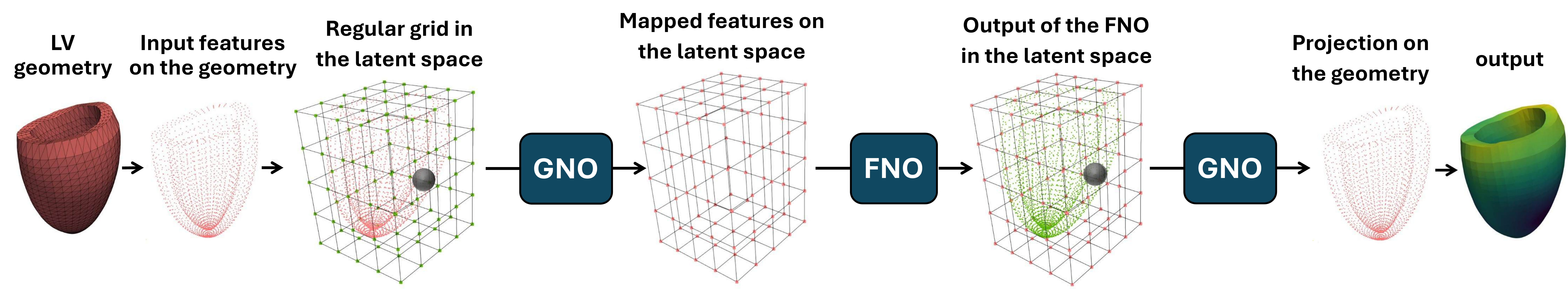}
    \caption{Schematic of the GINO architecture. The input LV geometry, discretized into a point cloud, is mapped onto a regular latent grid using the GNO encoder. For each node in the latent grid, a ball with radius $r$ aggregates features from points of the original geometry falling within this radius. This latent representation is subsequently processed by an FNO model. Finally, the resulting latent features are interpolated back onto the original geometry via the GNO decoder, where, for each original geometry point, features are gathered from adjacent nodes within a radius $r$ in the latent space.}
    \label{fig:GINO-Arch}
\end{figure}

\paragraph{GINO model.} The GINO architecture adopted in this study is based on the formulation by Li \textit{et al.}~\autocite{li2023geometryinformed} (\autoref{fig:GINO-Arch}). Neural operators generalize traditional feed-forward neural networks (FNNs), transitioning from mappings between finite-dimensional vectors to mappings between infinite-dimensional function spaces. Similar to FNNs, which consist of sequential linear transformations (e.g., matrix multiplications or convolutions) followed by pointwise nonlinear activations, neural operators incorporate sequences of linear operators combined with pointwise nonlinearities.

To formalize the neural operator, we consider an operator
\begin{align}
    \Psi: \mathcal{A} \to \mathcal{U},
\end{align}
mapping between function spaces $\mathcal{A}$ and $\mathcal{U}$. The neural operator $\Psi_\theta$ is constructed to estimate $\Psi$ with $L + 2$ layers, expressed as 
\begin{equation}
    \Psi_\theta := \mathcal{G}_{L+1} \circ \mathcal{G}_{L} \circ \cdots \circ \mathcal{G}_1 \circ \mathcal{G}_{0},
\end{equation}
where $\mathcal{G}_{0}$ is the initial lifting operator mapping inputs to a higher-dimensional space, and $\mathcal{G}_{L + 1}$ is the projection operator mapping to the output space. Both operators act pointwise and are typically parameterized by MLPs.

Each intermediate layer $\mathcal{G}_{\ell}: v_{\ell-1}(x) \to v_{\ell}(x)$ (for $\ell = 1, 2, \cdots, L$) is described by
\begin{equation}
    v_{\ell}(x) = \sigma\Big(W_{\ell} v_{\ell-1}(x) + \mathcal{K}_{\ell} \left(v_{\ell-1}\right)(x) + b_{\ell}\Big),
\end{equation}
where $\sigma$ denotes a pointwise nonlinear activation, $W_{\ell}$ and $b_{\ell}$ are the weight and the bias parameters, which are set to be constant functions here as in standard neural networks. We note $W_{\ell}$ and $b_{\ell}$ can also be arbitrary functions (e.g., MLPs). $\mathcal{K}_{\ell}$ is a linear operator differentiating neural operators from FNNs, enabling mappings between function spaces and is represented in integral form
\begin{equation} \label{eqn:kernel_function_1}
    \mathcal{K}_{\ell} \left(v_{\ell-1}\right)(x) = \int_D \kappa_{\ell}\left(x, y\right) v_{\ell-1}(y)dy,
\end{equation}
where $D$ is the domain of $v_{\ell-1}$, and $\kappa_{\ell}$ is a kernel function. To better capture nonlinear interactions, the kernel function $\kappa_{\ell}$ in \autoref{eqn:kernel_function_1} can be extended to depend on the input function $v_{\ell-1}$ as follows
\begin{equation} \label{eqn:kernel_function}
    \mathcal{K}_{\ell} \left(v_{\ell-1}\right)(x) = \int_D \kappa_{\ell}\Big(x, y, v_{\ell-1}(x), v_{\ell-1}(y)\Big)v_{\ell-1}(y)dy.
\end{equation}

The kernel function $\kappa_{\ell}$ is parameterized and learned during training along with the weights ($W_{\ell}$) and biases ($b_{\ell}$). Two prominent approaches are:
\begin{itemize}
    \item Graph neural operator (GNO): GNO~\autocite{GKNforPDEs2020}, localizes \autoref{eqn:kernel_function} by truncating the domain to a ball of radius $r$ around each point $x$
    \begin{align}
        \mathcal{K}_{\ell} \left(v_{\ell-1}\right)(x) = \int_{B_r(x)} \kappa_{\ell}\Big(x, y, v_{\ell-1}(x), v_{\ell-1}(y)\Big)v_{\ell-1}(y)dy.
    \end{align}
    Discretizing this integral via a Nystr\"{o}m approximation yields
    \begin{align} \label{eqn:GNO}
        \mathcal{K}_{\ell} \left(v_{\ell-1}\right)(x) \approx \sum_{i = 1}^{M} \kappa_{\ell}\Big(x, y_i, v_{\ell-1}(x), v_{\ell-1}(y_i)\Big)v_{\ell-1}(y_i) \mu(y_i),
    \end{align}
    where $\{y_1, y_2, \dots, y_M\} \subset B_r(x)$ and $\mu(y_i)$ are Riemannian sum weights. This formulation corresponds to a message-passing framework on graphs~\autocite{gilmer2017neural}.

    \item Fourier neural operator (FNO): FNO simplifies $\kappa_{\ell}\Big(x, y, v_{\ell-1}(x), v_{\ell-1}(y)\Big) := \kappa_{\ell}(x-y)$, and by applying the convolution theorem, the integral operator in \autoref{eqn:kernel_function} can be computed efficiently in the Fourier domain as ~\autocite{li2021Fourier}
    \begin{equation}
        \mathcal{K}_{\ell} \left(v_{\ell-1}\right)(x) = \mathcal{F}^{-1}\Big(R_{\ell} \cdot \mathcal{F}(v_{\ell-1})\Big)(x),
    \end{equation}
    where $\mathcal{F}$ and $\mathcal{F}^{-1}$ denote the Fourier and inverse Fourier transforms, respectively, and $R_{\ell}$ is the Fourier representation of the kernel $\kappa_{\ell}$. $\kappa_{\ell}$ is assumed to be periodic (by padding the input with zeros) to enable working with the discrete Fourier modes. In practice, when the input grid is uniform, the discrete Fourier transform (computed via fast Fourier transform~\autocite{AlgorithmforComplexFourier}) is employed.
\end{itemize}

GNO handles arbitrary geometries but may struggle with problems that involve long-range dependencies~\autocite{multipoleNO2020advances}, whereas FNO captures global interactions but is computationally efficient only on regular grids~\autocite{li2021Fourier}. GINO captures the advantages of both through the integration of three main components:
\begin{itemize}
    \item GNO encoder: The GNO encoder maps input LV point cloud features onto a latent regular grid via local integrals discretized as
    \begin{align} \label{eqn:GINO-encoder}
        \msfsl{a}(x^{\mathrm{grid}}) \approx \sum_{i = 1}^{M} \kappa\Big(x^{\mathrm{grid}}, y_i^{\mathrm{in}}\Big) \msfsl{a}\left(y_i^{\mathrm{in}}\right) \mu\left(y_i^{\mathrm{in}}\right),
    \end{align}
    where the integration is performed over the ball $B_{r_{\mathrm{in}}}(x^{\mathrm{grid}})$. The Riemannian weights $\mu(y_i^{\mathrm{in}})$ are computed to account for the local density of points.

    \item FNO core: The FNO model processes latent representations on the regular grid $\msfsl{a}(x^{\mathrm{grid}})$ efficiently capturing global interactions by operating in Fourier space.

    \item GNO decoder: The GNO decoder interpolates the latent grid back to the original LV geometry via local integrals discretized as
    \begin{equation} \label{eqn:GINO-decoder}
        u(x^{\mathrm{out}}) \approx \sum_{i = 1}^{S} \kappa\Big(x^{\mathrm{out}}, y_i^{\mathrm{grid}}\Big) u\left(y_i^{\mathrm{grid}}\right) \mu\left(y_i^{\mathrm{grid}}\right), 
    \end{equation}
    where $\{y_1^{\mathrm{grid}}, y_2^{\mathrm{grid}}, \dots, y_S^{\mathrm{grid}}\} \subset B_{r_{\mathrm{out}}}(x^{\mathrm{out}})$. In this step, since the latent grid is regular, the Riemannian weights are set uniformly.
\end{itemize}

Hyperparameters in the GINO model were prescribed following~\autocite{li2023geometryinformed}, with some adjustments. Specifically, we set the number of Fourier modes per spatial direction to 16 with a latent grid resolution of $28^3$. To reduce the number of trainable parameters and achieve parity with the GNN model, we applied a low-rank tensor factorization of the Fourier weights~\autocite{kossaifi2024multigrid} with rank = 0.01. The local integration radius was fixed at $r = 5$ mm yielding approximately 200{,}000 connectivity sizes per case, comparable to the GNN model.

\paragraph{Training and testing.} The 35{,}000 simulation cases were split into three sets: one with 17{,}500 cases for training and two with 8{,}750 cases each for development (hyperparameter optimization, abbreviated as ``dev'') and testing. Additionally, test cases on RR1 and RR2 were used to assess the discretization invariance of the trained models. YHC and HFC datasets were used to evaluate model generalizability on real-world LV geometries.

To quantify model performance at both regional (pointwise) and aggregate (subject-level) scales, two metrics were employed. The local absolute error at point $i$ is defined as
\begin{align} \label{eqn:local_error}
    e_i = \big|\hat{t}_{\mathrm{act}} - t_{\mathrm{act}}\big|_i,
\end{align}
where $\hat{t}_{\mathrm{act}}$ is the model prediction, and $t_{\mathrm{act}}$ is the ground truth from FE simulation. For each case $k$, the global relative error is calculated as
\begin{align} \label{eqn:global_error}
    E_k = \left(\dfrac{\dfrac{1}{n} \sum_{i=1}^{n} \big|\hat{t}_{\mathrm{act}} - t_{\mathrm{act}} \big|_{i}}{\dfrac{1}{n} \sum_{i=1}^{n} \left(t_{\mathrm{act}} \right)_{i}} \right)_{k},
\end{align}
where $n$ denotes the total number of sampled points in case $k$. The loss for a batch $B$ is defined by the mean global error
\begin{align} \label{eqn:loss}
    \mathcal{L}_B = \dfrac{1}{|B|}\sum_{k=1}^{|B|} E_k .
\end{align}

We added mini-batch support to the reference implementations of both models and for each model, chose the largest batch size $|B|$ that fits the GPU memory. To evaluate model robustness against input noise, Gaussian perturbations at levels of $1\%$, $5\%$, $10\%$, and $20\%$ of the channel-wise standard deviation were added to the coordinates $(x, y, z)$ and fiber directions $(\mathrm{f}_{0_{x}}, \mathrm{f}_{0_{y}}, \mathrm{f}_{0_{z}})$ in the test dataset.

The models were developed using \emph{PyTorch}~\autocite{PyTorch}, \emph{Torch Geometric}~\autocite{PyG2019}, and \emph{Neural Operator}~\autocite{kossaifi2024neural,kovachki2023neural}. Both GNN and GINO models were trained using the AdamW optimizer~\autocite{adamw2019} with an initial learning rate of $10^{-3}$, and a \texttt{ReduceLROnPlateau} learning rate scheduler that halves the learning rate after 50 stagnant epochs, down to a minimum of $10^{-6}$, after which training was stopped if no improvement in the training loss was observed for an additional 50 consecutive epochs. Mixed-precision computation (float16/float32) was used throughout, with Fourier operations in GINO restricted to float32 for numerical stability. The model achieving the lowest dev loss was selected for testing.

\subsection{Clinical application in CRT planning}
As an exploratory proof-of-concept illustrating how a fast surrogate could be embedded in a CRT optimization workflow, in a synthetic setting, we illustrated a potential post-training application of this study in CRT planning with a baseline case that had one intrinsic pacing site and a prolonged maximum activation time (a pathologic scenario associated with LBBB). Given the LV geometry and a noisy activation map (simulated and perturbed to mimic measurement noise in clinical maps), we first estimated the isotropic conductivity $d_{\mathrm{iso}}^{\ast}$ and the intrinsic pacing site $p_{\text{pacing}_1}^{\ast}$ by solving
\begin{align} \label{eqn:inv_prob}
    \left(d_{\mathrm{iso}}^{\ast},\; p_{\text{pacing}_1}^{\ast}\right) = \argmin_{\left(d_{\mathrm{iso}},\; p_{\text{pacing}_1}\right) } \left(E_k\right).
\end{align}

In this workflow, endocardial and epicardial fiber angles were set at $60^\circ$ and $-60^\circ$, respectively, to ensure the optimization tractable. The simulated activation time map was perturbed with spatially correlated Gaussian noise with mean zero, an empirical standard deviation of $\approx 5$~ms, extrema of $\approx \pm 20$~ms, and an effective spatial correlation length of $\approx 1$~cm on the LV surface. With $d_{\mathrm{iso}}^{\ast}$ and $p_{\text{pacing}_1}^{\ast}$ fixed, we then searched for an optimal lead position on the LV $p_{\text{pacing}_2}^{\ast}$ to minimize the maximum activation time,
\begin{align} \label{eqn:crt_opt}
    p_{\text{pacing}_2}^{\ast} = \argmin_{p_{\text{pacing}_2}} \Big(\max\big(t_{\mathrm{act}}\left(\mathbf{x}\right)\big)\Big).
\end{align}

Reduction in maximum activation time has been shown in prior studies to correlate with improved CRT response~\autocite{Wilczek2023, Martinez2023, Sweeney2010}. Both optimization problems (Eqs.~\ref{eqn:inv_prob}-\ref{eqn:crt_opt}) were solved using the differential evolution method~\autocite{Storn1997, Price2005} as implemented in \textit{SciPy}~\autocite{2020SciPy-NMeth}. This evolutionary approach was selected over the gradient-based methods, which are prone to converging to local optima.

\section*{Data availability}
The data is available at \url{https://zenodo.org/records/17651628}.

\section*{Code availability}
The code is available at \url{https://github.com/ehsanngh/DeepCardioSim}.

\section*{Acknowledgements}
This work was supported by NIH R01HL166508, R43HL174177 and NSF~2406029. This work used computational resources provided by the Institute for Cyber-Enabled Research (ICER) at Michigan State University.

\section*{Author contributions}
EN and LL contributed to the conception and design of the study. EN contributed to the code development, training of the model and wrote the initial draft of the manuscript. EN, HW, VZ, JG, GK, VB, SB and LL contributed to the analysis and interpretation of data, the editing and revising of  manuscript critically for important intellectual content, and final approval of the manuscript. All authors have significantly contributed to the submitted work and have reviewed the manuscript.

\section*{Declaration of competing interests}
The authors declare that they have no known competing financial interests or personal relationships that could have appeared to influence the work reported in this paper.

\printbibliography

\clearpage

\setcounter{figure}{0}
\renewcommand{\thefigure}{S\arabic{figure}}
\setcounter{table}{0}
\renewcommand{\thetable}{S\arabic{table}}

\section*{Supplementary information}
\subsection*{Details on generation of synthetic LV geometries} \label{appendix:geom_gen}
\begin{figure}[ht]
    \centering
    \includegraphics[width=0.48\textwidth]{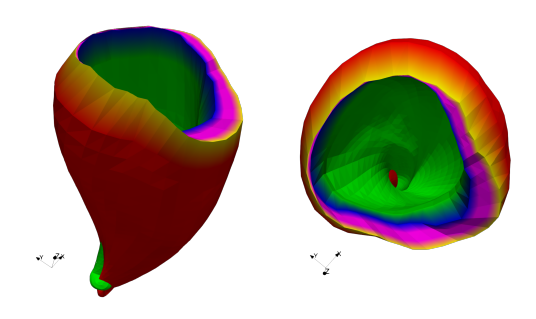}
    \caption{An example of an invalid generated synthetic LV geometry.}
    \label{fig:problematic_geom}
\end{figure}

To ensure the geometric validity of the generated LV meshes, we evaluated the element quality of each case using the scaled Jacobian metric. A mesh was considered problematic if its minimum scaled Jacobian value was below $10^{-4}$. In such cases, geometric distortions were typically localized near the apex, often manifesting as unrealistic twisting (see \autoref{fig:problematic_geom}).

For cases where the distortion was moderate, we attempted a localized mesh correction. Specifically, nodal positions within up to three apical layers were adjusted by interpolating their coordinates from neighboring nodes on adjacent layers, until the minimum scaled Jacobian exceeded $10^{-4}$. Geometries successfully corrected by this procedure were retained; otherwise, they were excluded from the dataset.

\begin{figure}[ht]
    \centering
    \includegraphics[width=0.96\textwidth]{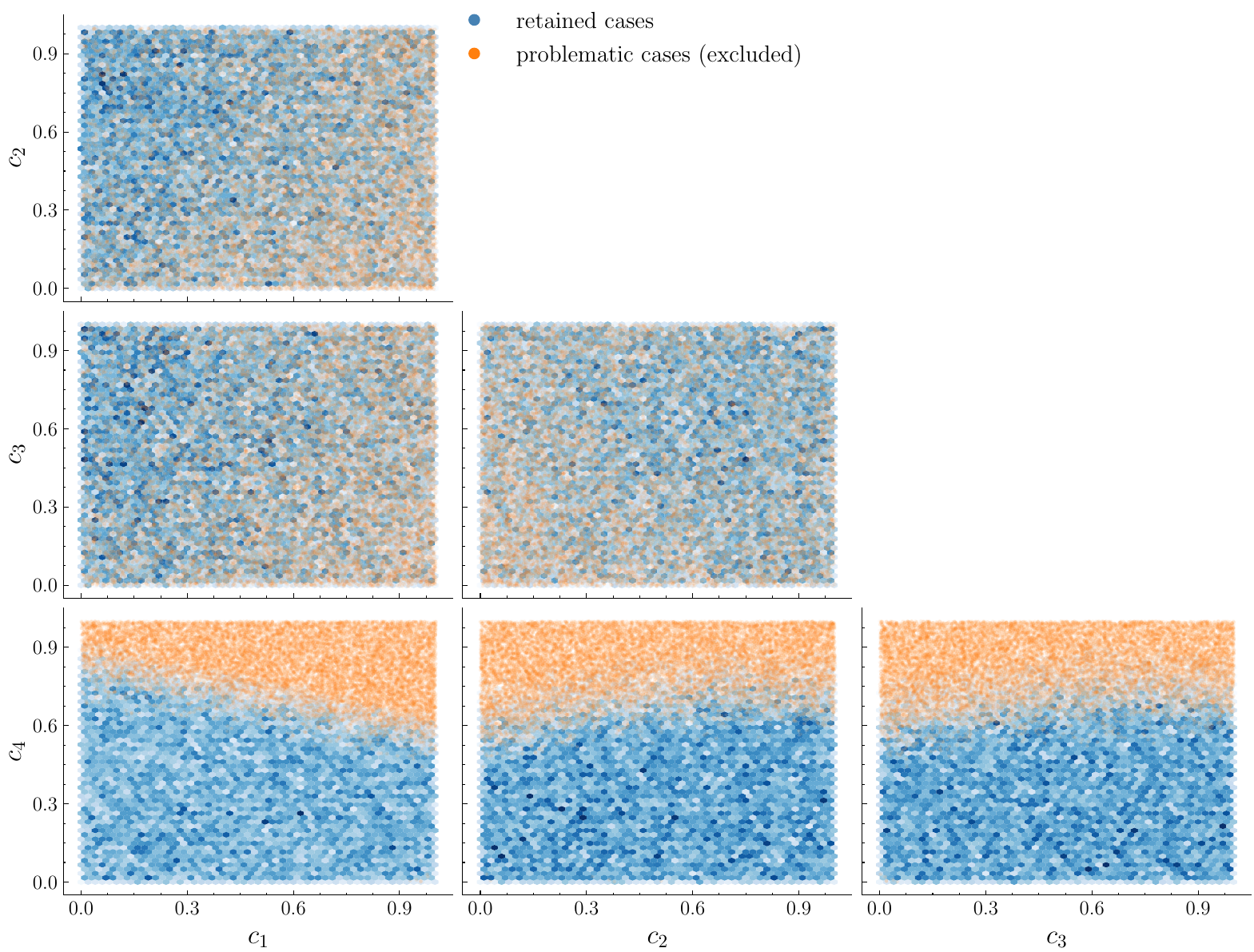}
    \caption{Distribution of POD coefficients for all generated LV geometries. Each point represents a synthetic LV geometry in the 4D POD coefficient space ($c_1$ - $c_4$). Retained cases (blue) passed the mesh quality check either initially or after mesh correction, whereas problematic cases (green) were excluded due to a minimum scaled Jacobian below $10^{-4}$. The large region of excluded geometries at large absolute values of the fourth POD coefficient suggests that the prescribed range for the fourth mode may require adjustment.}
    \label{fig:geom_coeff_dist}
\end{figure}

Out of an initial attempt to generate 50{,}000 synthetic LVs, approximately 15{,}000 geometries were excluded. The correction procedure successfully recovered roughly one in five problematic cases, where the distortion was less severe. As shown in \autoref{fig:geom_coeff_dist}, all problematic cases corresponded to large absolute values of the fourth POD coefficient, suggesting that the range prescribed for this mode in the original study~\autocite{BUOSO2021102066} may require adjustment when only four modes are used.

\subsection*{Mesh independence and time step sensitivity studies} \label{appendix:mesh_study}
The LV geometries generated by the shape model proposed by Buoso \textit{et al.} comprise 21{,}240 tetrahedral elements and 4{,}804 nodes. We explored two approaches to refine these meshes.

In the first approach, each element was subdivided into eight smaller tetrahedra, resulting in 169{,}920 elements after the first refinement and 1{,}359{,}360 elements after the second refinement. This procedure could be repeated once more to obtain an even finer discretization.

In the second approach, we adopted the base mapping methodology used in Buoso \textit{et al.}'s study, which maps a reference tetrahedral mesh defined on a unit disc onto the LV anatomy. This method provides direct control over the number of elements along the longitudinal, radial, and transmural directions, unlike the uniform subdivision of the first approach. The same mapping procedure was also used to convert the surface meshes of real LV geometries from the online dataset~\autocite{ChoKim2021} into tetrahedral meshes consistent with those used for the synthetic geometries in training.

We observed that the second approach performs best when starting from a fine mesh and subsequently coarsening it, whereas attempting to refine a coarse mesh using this method often results in poor element quality due to interpolation artifacts. Therefore, we adopted a combined strategy: fine meshes were first generated using the subdivision approach, then systematically coarsened using the mapping-based approach to achieve finer control over the final element count.

\begin{figure}[ht]
    \centering
    \includegraphics[width=0.96\textwidth]{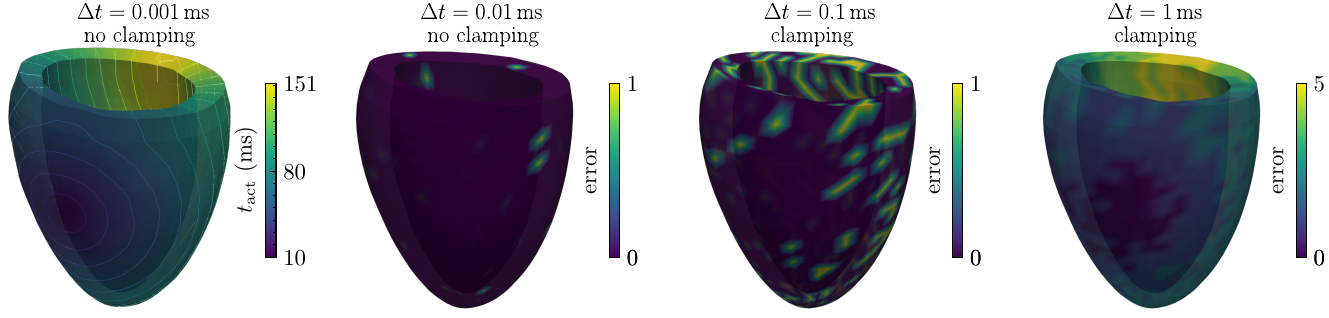}
    \caption{Time step sensitivity study. Activation time map at $\Delta t = 0.001$~ms (reference) and difference maps with respect to the reference for $\Delta t = 0.01$, $0.1$, and $1$~ms.}
    \label{fig:time_ind}
\end{figure}

Using the mesh obtained by subdividing the base mesh twice (1.36M elements), we solved the governing equations using time steps $\Delta t \in \{0.001, 0.01, 0.1, 1\}$~ms. To improve numerical robustness, the pacing stimulus was applied with a smooth ramp over a 2~ms interval and within a radius of 0.5~mm around the pacing point, rather than as an instantaneous regional stimulus. In addition, we applied a clamping safeguard to the recovery variable $\eta$, enforcing $\eta \ge -0.1$, to prevent nonphysical values and improve stability for larger time steps. This safeguard remained inactive for $\Delta t = 0.001$ and $0.01$~ms, but was occasionally activated for $\Delta t = 0.1$ and $1$~ms during a cardiac cycle. As shown in \autoref{tab:time_ind} and \autoref{fig:time_ind}, the activation time map for $\Delta t = 0.1$ differs from the reference solution ($\Delta t = 0.001$) by at most 1~ms. For $\Delta t = 1$~ms, the maximum difference increases to 5~ms. Nevertheless, given the substantial computational savings and that activation time maps typically carry a few-millisecond uncertainty in experimental and clinical settings~\autocite{Narayan2024}, we used $\Delta t = 1$~ms in the subsequent mesh resolution analysis.

\begin{table}[ht]
    \caption{Time step sensitivity study.}
    \label{tab:time_ind}
    \centering
    \begin{tabular}{cccc}
    \toprule
    \multirow{2}{*}{\begin{tabular}[c]{@{}c@{}}$\Delta t$ \\(ms)\end{tabular}} & \multirow{2}{*}{\begin{tabular}[c]{@{}c@{}}$\max(t_{\mathrm{act}})$ \\(ms)\end{tabular}} & \multirow{2}{*}{\begin{tabular}[c]{@{}c@{}}$\max(\mathrm{error})$ \\(ms)\end{tabular}} & \multirow{2}{*}{\begin{tabular}[c]{@{}c@{}}clamping \\activated\end{tabular}} \\
        & & & \\
    \midrule
    0.001 & 151  & - & no  \\
    0.01 & 151 & 1 & no \\
    0.1 & 152 & 1 & yes \\
    1 & 156 & 5 & yes \\
    \bottomrule
    \end{tabular}
\end{table}

The FE simulations were then tested on six meshes, ranging from 21k to 4.4M elements, and for two time step values, $\Delta t = 1$~ms and $\Delta t = 0.1$~ms.

\begin{figure}[ht]
    \centering
    \includegraphics[width=0.45\textwidth]{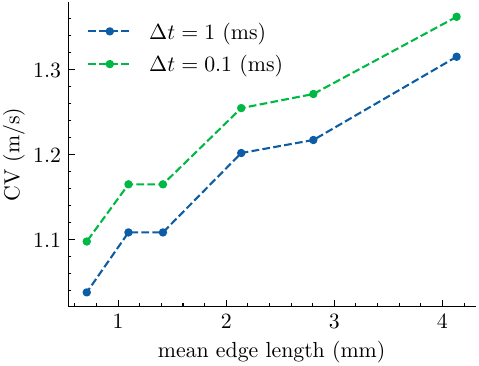}
    \caption{Mesh-resolution study of LV activation time.}
    \label{fig:mesh-ind}
\end{figure}

\autoref{tab:mesh_ind} and \autoref{fig:mesh-ind} summarize the results of the mesh independence study for a representative LV geometry. The pacing-site location and the position of the point with maximum activation time remained consistent across all simulations, with a geodesic distance of $11.42 \pm 0.03$~cm between them. The conduction velocity (CV) was computed by dividing this distance by the activation time difference between these two points, neglecting fiber orientation for simplicity.

\begin{table}[ht]
    \caption{Mesh independence study.}
    \label{tab:mesh_ind}
    \centering
    \begin{tabular}{ccccccc}
    \toprule
    \multirow{3}{*}{\# nodes} & \multirow{3}{*}{\# cells} & \multirow{3}{*}{\begin{tabular}[c]{@{}c@{}}mean edge \\length \\(mm)\end{tabular}} & \multicolumn{2}{c}{$\Delta t = 1$ (ms)} & \multicolumn{2}{c}{$\Delta t = 0.1$ (ms)} \\
    \cmidrule(lr){4-5}
    \cmidrule(lr){6-7}
     & & & $\max(t_{\mathrm{act}})$ & $CV$ & $\max(t_{\mathrm{act}})$ & $CV$ \\
     & & & (ms) & $(\textrm{m}/\textrm{s})$ & (ms) & $(\textrm{m}/\textrm{s})$ \\ 
    \midrule
    4804 & 21,240  & 4.13 & 92  & 1.31 & 89 & 1.36  \\
    14,970 & 73,950 & 2.81 & 99  & 1.22 & 95 & 1.27 \\
    33,327 & 169,920 & 2.14 & 100 & 1.20 & 96 & 1.25 \\
    103,849 & 557,550 & 1.41 & 108 & 1.11 & 103 & 1.16 \\
    246,493 & 1,359,360  & 1.10 & 108  & 1.11 & 103 & 1.16  \\
    781,394 & 4,428,810  & 0.71 & 115 & 1.04 & 109 & 1.10 \\
    \bottomrule
    \end{tabular}
\end{table}

\begin{figure}[ht]
    \centering
    \includegraphics[width=0.9\textwidth]{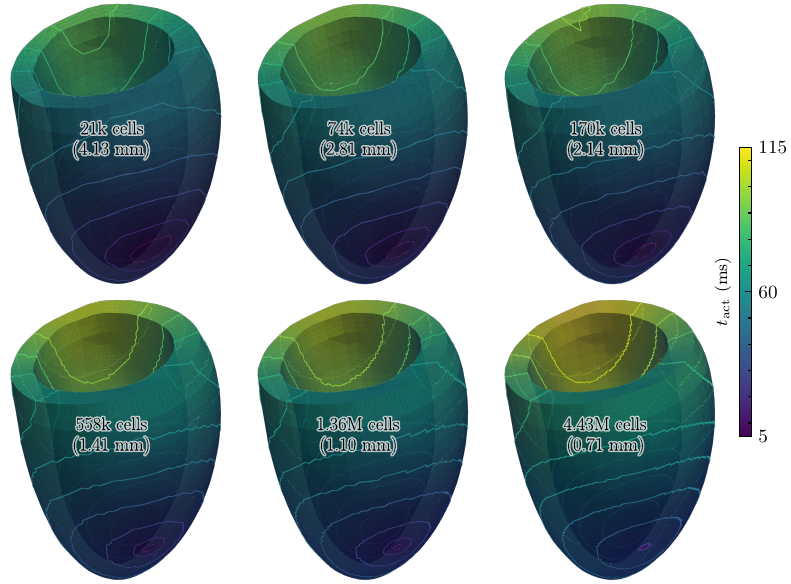}
    \caption{Mesh-resolution study of LV activation time for $\Delta t = 1$~ms. Six tetrahedral meshes, ordered from coarse to fine (21k-4.43M cells), are shown; mean edge lengths are reported in parentheses.}
    \label{fig:mesh-study}
\end{figure}

\autoref{fig:mesh-study} illustrates the results for $\Delta t = 1$~ms. Although the solutions did not fully converge with further mesh refinement (partly because the localized pacing effect diminishes with decreasing element size, slightly delaying the maximum activation time), the overall activation patterns and trends remained consistent. Since our analysis focuses primarily on the pacing-induced reduction in activation time rather than its absolute magnitude, we selected the 1.3M-element mesh with $\Delta t = 1$~ms as an optimal compromise between accuracy and computational efficiency for dataset generation.

\subsection*{Pilot hyperparameter optimization}
\autoref{tab:param_opt} summarizes a pilot hyperparameter study for the GNN and GINO models conducted on a previous version of the synthetic dataset. For the GNN, we varied the embedding dimension $\msfsl{d}_{\msfsl{emb}} \in \{16, 32, 64, 128\}$. For GINO, we varied the number of Fourier modes per spatial direction $\msfsl{m} \in \{4, 8, 16, 32\}$ and used a low-rank factorization with rank $=0.4$ (in the main study, rank $=0.01$ was used to keep the number of trainable parameters comparable to the GNN). Overall, this pilot study suggests that increasing GINO capacity can reduce error on the synthetic dataset, whereas increasing the GNN embedding dimension does not consistently improve performance under the tested settings. Because the dataset and training protocol differ from the main study, these results are provided for context and are not directly comparable to the metrics reported in the main paper.

\begin{table}[ht]
    \caption{Pilot hyperparameter study on a previous version of the synthetic dataset. GNN models with different embedding dimensions $\msfsl{d}_{\msfsl{emb}}$ and GINO models with different numbers of Fourier modes per spatial direction $\msfsl{m}$. \(\mathcal{L}\) is the mean relative error (\autoref{eqn:loss}) over each set.}
    \label{tab:param_opt}
    \centering
    \begin{tabular}{ccccc}
    \toprule
    \multirow{2}{*}{model} & \multirow{2}{*}{\begin{tabular}[c]{@{}c@{}}GNN: $\msfsl{d}_{\msfsl{emb}}$ \\GINO: $\msfsl{m}$\end{tabular}} & \multirow{2}{*}{\begin{tabular}[c]{@{}c@{}}\# params \\(million)\end{tabular}} & \multicolumn{2}{c}{$\mathcal{L}$} \\
    \cmidrule(lr){4-5}
     & & & train & dev \\
    \midrule
    \multirow{4}{*}{GNN} & 16 & 0.63 & 0.0414  & 0.0431 \\
    & 32 & 1.29 & 0.0296 & 0.0334 \\
    & 64 & 3.93 & 0.0265 & 0.0317 \\
    & 128 & 14.4 & 0.0267 & 0.0326 \\
    \midrule
    \multirow{4}{*}{\begin{tabular}[c]{@{}c@{}}GINO \\(rank = 0.4)\end{tabular}} & 4  & 0.97  & 0.0130 & 0.0142 \\
    & 8  & 8.64  & 0.0098 & 0.0117 \\
    & 16 & 66.7 & 0.0081 & 0.0116 \\
    & 32 & 476.7 & 0.0080 & 0.0114 \\
    \bottomrule
    \end{tabular}
\end{table}

\subsection*{Additional results} \label{appendix:sup_res}
\begin{figure}[ht]
    \centering
    \includegraphics[width=0.96\textwidth]{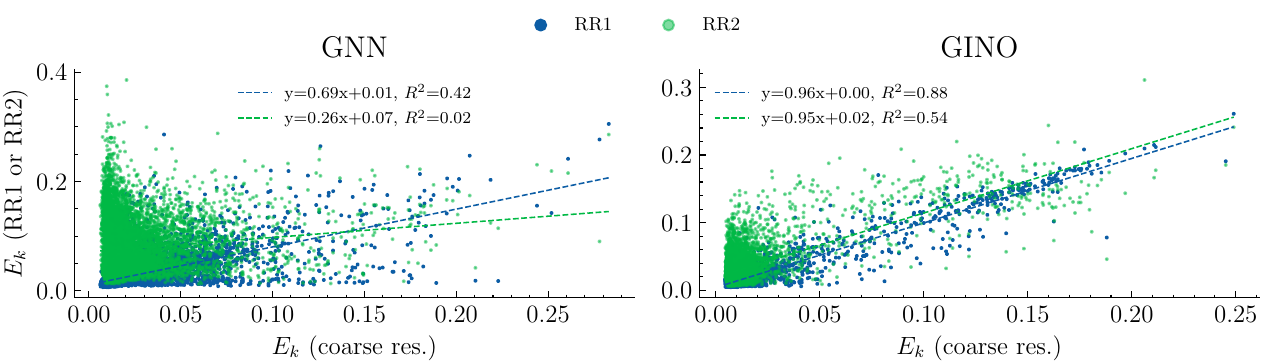}
    \caption{Relationships between prediction errors on the coarse-resolution mesh (horizontal axis) and the refined meshes RR1 and RR2 (vertical axis) for the GNN and GINO models. Each point corresponds to one test set case. Linear regression fits are shown with their corresponding $R^2$ values. The GINO model exhibits strong (RR1) to moderate (RR2) correlation across resolutions, whereas the GNN model shows weaker correlation, particularly on RR2.}
    \label{fig:baseres_VS_RR1RR2}
\end{figure}

\autoref{fig:baseres_VS_RR1RR2} examines how prediction errors on the coarse resolution of the test dataset (4{,}804 points) relate to those on the refined discretizations RR1 (33{,}327 points) and RR2 (246{,}493 points). For RR1, both models exhibit a predominantly linear relationship between coarse and refined resolution errors, although the correlation is substantially stronger for the GINO. On RR2, the GINO model preserves a moderate linear trend, whereas the GNN shows almost no correspondence between errors at the base resolution and those on RR2.

\begin{figure}[htp]
\centering
\includegraphics[width=\textwidth]{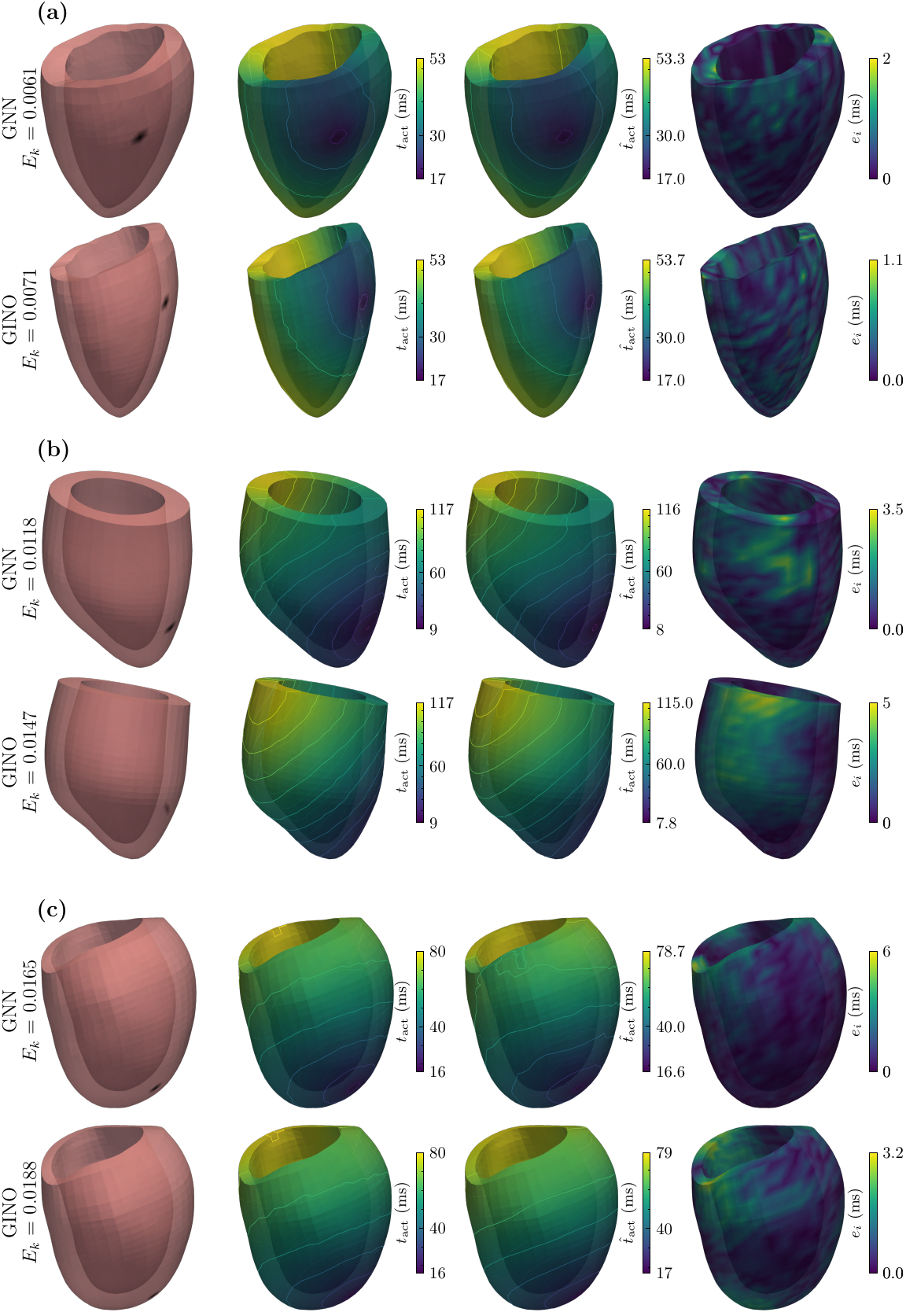}
\caption{Examples with global error $E_k$ smaller than the dataset mean. (a) Test dataset. (b) YHC dataset. (c) HFC dataset. In each dataset, a single case with $E_k$ lay below the mean for both models (GNN and GINO) was selected. Refer to \autoref{fig:sample_results_mean} for organization.}
\label{fig:sample_results_below}
\end{figure}

\begin{figure}[htp]
\centering
\includegraphics[width=\textwidth]{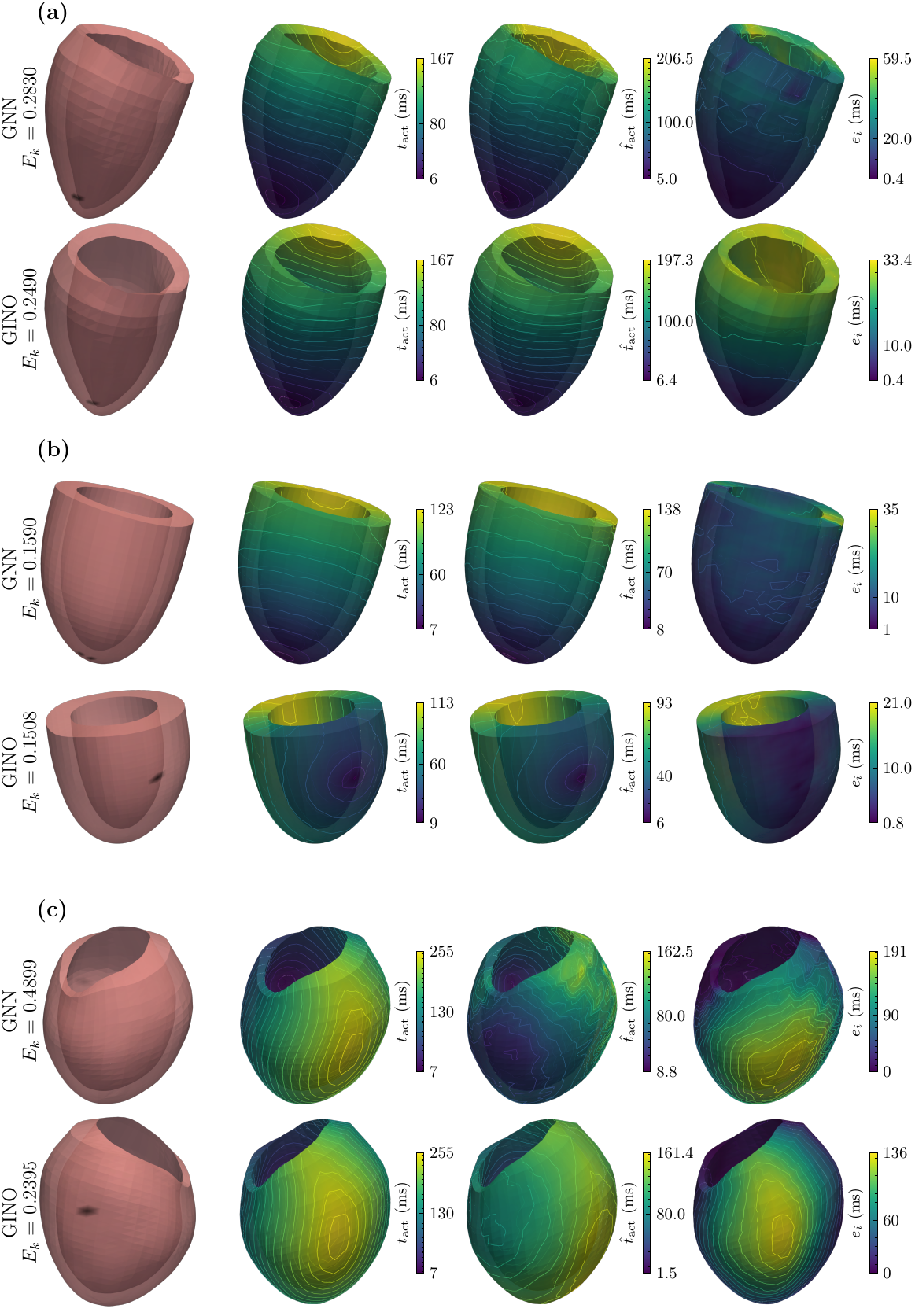}
\caption{Cases with the maximum global error $E_k$ in each dataset. Refer to \autoref{fig:sample_results_mean} for organization. For (a) and (c), the same case showed maximum error for both models.}
\label{fig:sample_results_max}
\end{figure}

Figs.~\ref{fig:sample_results_below} and \ref{fig:sample_results_max} present additional representative cases from the test, YHC, and HFC datasets.

\end{document}